\documentclass[acmsmall]{acmart}
\acmJournal{TOPLAS}
\acmVolume{1}
\acmNumber{1} % CONF = POPL or ICFP or OOPSLA
\acmArticle{1}
\acmYear{2022}
\acmMonth{1}
\acmDOI{} % \acmDOI{10.1145/nnnnnnn.nnnnnnn}
\startPage{1}

\usepackage{booktabs}   %% For formal tables:
\usepackage[shortlabels]{enumitem}
\usepackage{subcaption} %% For complex figures with subfigures/subcaptions
\usepackage{subfiles}
\usepackage{color}
\usepackage{amsmath}
\usepackage{xspace}
\usepackage{cleveref}
\usepackage{listings} % for code
\usepackage{syntax} % for code highlighting
\usepackage{xcolor} % for code highlighting
\usepackage{graphicx}
\usepackage{cprotect}

\usepackage{pifont}
\newcommand{\cmark}{\ding{51}}%
\lstdefinelanguage{coq}{
    keywords={Repair, Module, End, module, Theorem, Proof, Record, Lemma, Definition, Abort, Qed, forall, Inductive, Type, Prop, Set, fun, fix, forall, Require, Import, Fixpoint, match, end, with, as, return, struct, Qed, Defined, let},
    basicstyle=\linespread{0.95}\small\ttfamily,
    keywordstyle=\color{blue},
    commentstyle=\itshape\rmfamily,
    showstringspaces=false,
    columns=flexible,
    breaklines=true,
    texcl=true,
    mathescape=true,
    tabsize=4,
    stringstyle=\color{brown},
    escapeinside={(@}{@)},
}

\newcommand{\toolname}{Passport\xspace}

\newcommand{\reducedstrut}{\vrule width 0pt height .9\ht\strutbox depth .9\dp\strutbox\relax} % for \codediff
\newcommand{\codesima}[1]{%
  \begingroup
  \setlength{\fboxsep}{0pt}%
  \colorbox{orange!25}{\reducedstrut\texttt{#1}\raisebox{0.8ex}{\scalebox{0.66}{2}}\/}%
  \endgroup
} % to highlight the similarities between two code blocks
\renewcommand{\textsuperscript}[1]{}
\newcommand{\codesimb}[1]{%
  \begingroup
  \setlength{\fboxsep}{0pt}%
  \colorbox{red!25}{\reducedstrut\texttt{#1}\raisebox{0.8ex}{\scalebox{0.66}{1}}\/}%
  \endgroup
} % to highlight the similarities between two code blocks
\newcommand{\codesimc}[1]{%
  \begingroup
  \setlength{\fboxsep}{0pt}%
  \colorbox{yellow!25}{\reducedstrut\texttt{#1}\raisebox{0.8ex}{\scalebox{0.66}{3}}\/}%
  \endgroup
} % to highlight the similarities between two code blocks
\newcommand{\va}[1]{\codesima{#1}}
\newcommand{\vb}[1]{\codesimb{#1}}
\newcommand{\vc}[1]{\codesimc{#1}}

\hypersetup{draft}

\setcopyright{none}
\begin{document}

\title[Passport: Improving Automated Formal Verification Using Identifiers]{Passport: Improving Automated Formal Verification Using Identifiers}
\author{Alex Sanchez-Stern\**}
\affiliation{
  \institution{University of Massachusetts Amherst}
  \country{USA}
}
\email{sanchezstern@cs.umass.edu}

\author{Emily First\**}
\affiliation{
  \institution{University of Massachusetts Amherst}
  \country{USA}
}
\email{efirst@cs.umass.edu}
\thanks{\** Co-first authors}

\author{Timothy Zhou}
\affiliation{
  \institution{University of Illinois Urbana-Champaign}
  \country{USA}
}
\email{ttz2@illinois.edu}

\author{Zhanna Kaufman}
\affiliation{
  \institution{University of Massachusetts Amherst}
  \country{USA}
}
\email{zhannakaufma@cs.umass.edu}

\author{Yuriy Brun}
\affiliation{
  \institution{University of Massachusetts Amherst}
  \country{USA}
}
\email{brun@cs.umass.edu}

\author{Talia Ringer}
\affiliation{
  \institution{University of Illinois Urbana-Champaign}
  \country{USA}
}
\email{tringer@illinois.edu}


\begin{abstract}

Formally verifying system properties is one of the most effective ways
of improving system quality, but its high manual effort requirements
often render it prohibitively expensive.
Tools that automate formal verification, by learning from proof
corpora to suggest proofs, have just begun to show their promise.
These tools are effective because of the richness of the data the proof
corpora contain.
This richness comes from the stylistic conventions followed by
communities of proof developers, together with the powerful logical
systems beneath proof assistants.
However, this richness remains underexploited, with most work thus
far focusing on architecture rather than on how to make the most of
the proof data.

In this paper, we develop \toolname, a fully-automated proof-synthesis tool
that systematically explores how to most effectively exploit one aspect of
that proof data: identifiers.
\toolname enriches a predictive Coq model used by proof-synthesis tools with
three new encoding mechanisms for identifiers: category vocabulary
indexing, subword sequence modeling, and path elaboration.
We compare \toolname to three existing base tools which \toolname
can enhance: ASTactic, Tac, and Tok.
In head-to-head comparisons, \toolname automatically proves 29\% more
theorems than the best-performing of these base tools.
Combining the three \toolname-enhanced tools automatically proves 38\%
more theorems than the three base tools together, without \toolname's
enhancements.
Finally, together, these base tools and \toolname tools enhanced
with identifier information prove 45\% more theorems than the combined
base tools without \toolname's enhancements.
Overall, our findings suggest that modeling identifiers can play a
significant role in improving proof synthesis, leading to
higher-quality software.

% Finally, we discuss our experiences building a machine learning tool for
% proof synthesis, as it compares to building traditional proof automation
% tools, in hopes to shine light on common challenges and potential solutions
% during a potentially pivotal moment in our field. \todo{Yuriy is not sure
% this paragraph rises to the level of
% abstract. Ironically, the paragraph is too abstract. I'd maybe drop it here.}

\end{abstract}

\maketitle

\section{Introduction}
\label{sec:introduction}

%Yuriy:
%
%This introduction is written a little too much toward an expert in not
%just proof assistants, but proof synthesis. We need to make it a little bit
%broader-reaching to accomodate some reviewers we are likely to get. These
%folks won't have a good intuition for what a proof script looks like, what
%identifiers are and why they're important. They'll just feel lost. A few
%relatively small, high-level explanations about the various terms we use, and
%some intuition in the intro, will go a long way.

Verifying software with proof assistants gives engineers the potential to prove the absence
of costly and possibly dangerous bugs, leading toward more reliable software systems.
Teams of specialized experts have already realized this potential for large
and critical systems, such as operating system microkernels~\cite{sel4},
distributed systems~\cite{verdi}, and compilers~\cite{compcert}, among
hundreds of other formally verified software systems~\cite{PGL-045}.
These advances have already had significant impact on industry. For example,
Airbus France uses the CompCert~\cite{compcert} C compiler to ensure safety and
improve performance~\cite{Souyris14};
Chrome and Android both use cryptographic code formally verified in Coq
to secure communication~\cite{Erbsen19}.
%; Mozilla has its own verified cryptography
%library for Firefox, which improves performance~\cite{Jacobs20}.
But the full potential of these proof assistants still remains far
from realized, as the costs of verified software development and
maintenance remain high, even for experts~\cite{replica}.

To prove theorems in these proof assistants, proof engineers typically
write high-level sequences of strategies called \emph{proof scripts},
which guide the proof assistant toward low-level, machine-checkable
representations called \emph{proof objects}~\cite{PGL-045}.
In recent years, techniques that use machine learning to synthesize
these proof scripts have shown promise in alleviating some of the effort of
verification~\cite{proverbot, First20, First22icse, Yang19, christian-gnn}.
These \emph{proof-synthesis} tools learn from corpora of existing proof
scripts and theorems to automate the construction of proof scripts for new theorems.
In particular, these tools build predictive models of proof
scripts, and then use search to explore the proof-script space.
This process uses the proof assistant to guide the search and evaluate
ultimate success.

In this paper, we explore ways of improving these predictive models by
better exploiting the richness of the proof data that they learn from.
We focus in particular on modeling \emph{identifiers}: the names that
uniquely identify theorems, datatypes, functions, type constructors,
and local variables.
Previous machine-learning-guided proof-synthesis tools have either
ignored the names of individual identifiers completely and only
encoded basic categorical information about them, or given common
identifiers unique indices and marked all others as unknown, without
category information.
We use our approach to build \toolname: a proof-synthesis tool for the
Coq proof assistant~\cite{coq} that enriches its models with three new
encoding mechanisms for these identifiers: category vocabulary
indexing, subword sequence modeling, and path elaboration.
We show that all three of these encodings improve performance of the
model.

%% \paragraph{Identifiers in Proof Synthesis}

%% The three encoding mechanisms that we introduce serve as new points in the existing design
%% space for encoding identifiers for proof-synthesis tools.
%% Some proof-sythesis tools~\cite{Yang19, First20, First22icse} completely
%% generalize identifiers,
%% giving every identifier the same \verb|ident|
%% token, and erasing its actual name. Other tools~\todo{cite some} hyper-specialize and treat
%% every identifier as unique. The former is useful for generalization,
%% but less useful for learning about particular identifiers or groups of identifiers;
%% the latter is useful for learning about particular identifiers, but less useful
%% for generalization.

%% To get the best of both worlds, some tools use a hybrid approach, treating
%% common identifiers as unique, with all other identifiers
%% represented by a single \verb|unknown-ident|
%% token~\cite{proverbot, christian-gnn}.
%% This allows for basic handling of out-of-vocabulary identifiers,
%% without sacrificing the ability to learn about common ones.
%% To the best of our knowledge, prior to our work with \toolname, this was the
%% state of the art of modeling identifiers in proof-synthesis tools.

\paragraph{Identifiers in \toolname}

%\todo{this first sentence needs to become more formal, identifying the
%problem unsolved the prior approaches.}

\toolname goes beyond existing techniques for proof synthesis by encoding
identifiers with \textit{three different encoding mechanisms} (described in
Sections~\ref{sec:overview}~and~\ref{sec:encodings}):

\begin{enumerate}
\item \textbf{Category Vocabulary Indexing}: \toolname encodes each
  identifier with the category it comes from (global definition, local
  variable, or type constructor); and for the most common identifiers
  in each category, \toolname encodes indices corresponding to their
  names. That is, each common identifier is given a unique tag,
  associating it with all other uses of that exact identifier.
\item \textbf{Subword Sequence Modeling}:
  For all identifiers, \toolname uses a subword sequence model to
  draw bridges between related names. That is, identifiers are broken
  down into common word-pieces, and processed with a sequence model.
\item \textbf{Path Elaboration}: For type constructors and global definitions,
  \toolname encodes their \emph{fully-qualified paths}---the names of directories,
  files, and modules within which they are contained.
\end{enumerate}
While we focus on Coq in this paper, similar techniques should apply for other proof
assistants, including Lean~\cite{lean}, Isabelle/HOL~\cite{isabelle}, and
Agda~\cite{agda}.

\paragraph{Results}

We evaluated \toolname on the CoqGym benchmark suite~\cite{Yang19}, a
common benchmark for proof-synthesis tools composed of 124 open-source
Coq projects.
We compare to three existing machine-learning-guided proof-synthesis
tools, ASTactic~\cite{Yang19}, Tac, and Tok~\cite{First20}.
We found that all three of our encoding mechanisms improve \toolname's
performance, in terms of being able to prove more theorems fully
automatically.
For example, adding path elaboration leads to proving 12.6\% more
theorems.
We also measured the impact of adding identifier information to each
of the categories of identifiers individually, and found that
\toolname's approach is useful for each.

Together with the three prior tools, \toolname is able to fully automatically
prove 1,820 of the 10,782 theorems in our benchmark test set, whereas without
\toolname, these prior tools combined can prove 1,259 theorems. That is an
increase of 45\% theorems proven over this prior work.

%% Furthermore, in line with existing work in program synthesis~\cite{big-code, fast-code},
%% we find that BPE tokenization further improves synthesis performance.
%% \todo{Summarize BPE evaluation.}
%% \todo{Summarize BPE results.}
% TODO whatever other eval Alex does here

\paragraph{Contributions}

The main contributions of our work are:
\begin{enumerate}

  \item New techniques (Section~\ref{sec:encodings}) for encoding
    identifiers in a proof assistant context, and \toolname, an
    implementation of these techniques within an existing
    proof-synthesis framework. \toolname is
    open-source.\footnote{\url{https://github.com/LASER-UMASS/Passport}}

  \item An evaluation (Section~\ref{sec:evaluation}) comparing
    \toolname with prior work ASTactic and TacTok, showing that
    (1)~each mechanism for encoding identifiers helps \toolname model
    proof scripts more precisely and improves performance of proof
    synthesis, and (2)~encoding each identifier category alone is
    still an improvement over not encoding any.

  \item A forward-looking discussion (Section~\ref{sec:discussion}) of
    the challenges that we faced when building \toolname (relative to
    building symbolic proof automation), along with potential
    solutions to those challenges. Crucially, our evaluation includes
    an experiment (Section~\ref{ssec:nondeterminism}) measuring the
    impact of non-deterministic training variance on our tool.

\end{enumerate}

% Talia: I hate outline paragraphs, but I'm open to them if someone really wants them
%The remainder of this paper is structured as follows:
%Section~\ref{sec:motivating} illustrates our approach with an example.
%\todo{\ldots}

\section{Background on proofs and proof synthesis}
\label{sec:background}

\begin{figure}[t]
  \includegraphics[width=0.6\columnwidth]{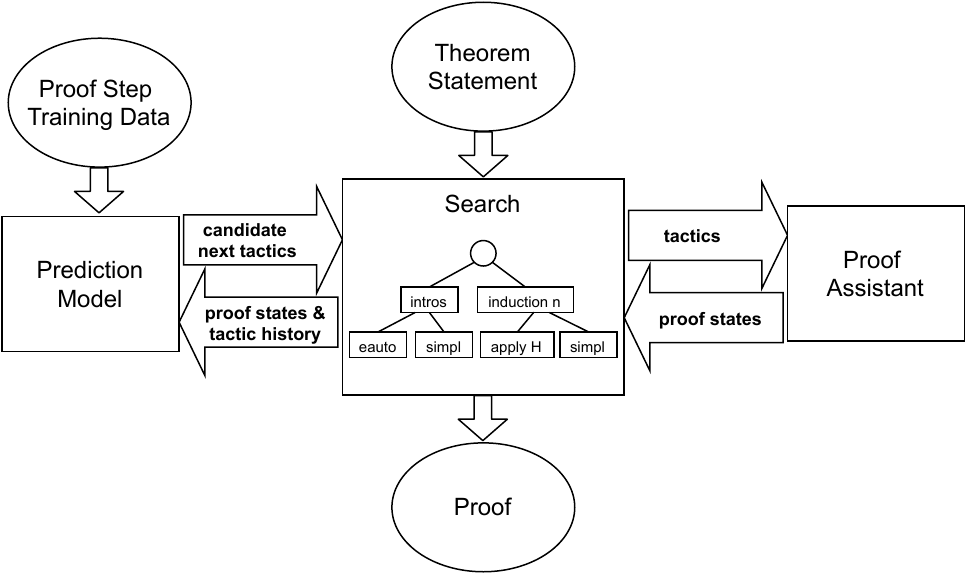}
  \caption{The system architecture of a
    machine-learning-prediction-guided proof-synthesis tool.}
  \label{fig:proof-synthesis-background}
% source: https://drive.google.com/file/d/1bqN96LiSWPOhyrXhOPiafYzJXZhBj0AL/view?usp=sharing
\end{figure}

To write proofs in Coq, the proof engineer starts by stating a theorem
to prove.
They then write a proof that this theorem holds.
Every theorem in Coq is a type definition, described in a rich type
system; writing a proof in Coq amounts to finding a term with the
stated theorem type.\footnote{This refers to the Curry-Howard
correspondence, which shows type systems and proof systems to be
equivalent.}

But doing this directly can be challenging, so instead, proof
engineers write these proofs interactively with Coq's help.
At each step, proof engineers pass Coq high-level strategies called
\emph{tactics}, and Coq responds with the current proof obligations
after executing each tactic.
Each tactic guides Coq in a search for a term with the stated type,
refining the state until no new obligations hold.
At that point, the proof engineer has written a sequence of tactics
called a \emph{proof script} (like the one in
Figure~\ref{fig:proof-script})---and Coq, for its part, has
constructed a \emph{proof term} or \emph{proof object} with the stated
type. The language of proof scripts in Coq is called Ltac, and the
language of proof terms in Coq, as well as programs and definitions,
is called Gallina.

In recent years, machine-learning-guided proof-synthesis tools have
been developed which aim to make the burden of proving easier by
automatically generating the proof script, instead of asking the user
to write it.
While the approaches of these tools can differ, most share similar
components and structure.
Figure~\ref{fig:proof-synthesis-background} shows the common
architecture of most machine-learning-guided proof-synthesis tools.

%% Automated proof-synthesis tools are tools that mimic the human
%% interaction with Coq to produce these \emph{proof scripts}.
%% %
At the heart of these tools is the prediction model, which guides the
proof search by producing \emph{tactic predictions}, or candidate next
tactics, at every step.
Every prediction model takes some set of information about the proof
state or proof script, and produces a set of candidate tactics.
Crucial to doing so is the ability to encode information about the
current state of the proof so far into feature vectors, which can be
used to train a tactic model.

\paragraph{ASTactic and TacTok}

\toolname's tactic model architecture inherits the design choices of
ASTactic~\cite{Yang19} for encoding the proof obligations and
TacTok~\cite{First20} for encoding the proof script.

Proof obligations consist of the goals to be proven, local context,
and the environment.
Each term of the proof state has an underlying abstract syntax tree
(AST) representation.
ASTactic serializes these ASTs and uses a TreeLSTM~\cite{Tai15} to
encode them~\cite{Yang19}. TacTok adopts this encoding for the proof
state.

The proof script consists of a sequence of tokens in Ltac.
Before encoding these tokens, each proof script is preprocessed to
remove high-frequency low-signal tokens, such as punctuation.
TacTok uses a Bidirectional LSTM~\cite{Peters18} to encode this
sequence of tokens~\cite{First20}.

The model of ASTactic and TacTok is trained using supervised learning
with a set of human-written proofs to predict the next proof step
(tactic and arguments) of an incomplete proof.
A limited generative tree-grammar tactic model, adopted from
ASTactic~\cite{Yang19}, makes these downstream predictions.
While there may be many valid proofs for a single theorem statement,
there is no clear way of determining how appropriate an alternative
tactic or proof is, so the model is taught to imitate human-written
proofs.

\section{Overview}
\label{sec:overview}

The proof state is made up of many Gallina terms; modeling these terms
well is key to producing accurate models.
However, previous models have left out much of the essential
information about identifiers in terms, when they have encoded
identifiers at all.
Encoding identifiers well is essential because proof corpora in Coq
are rich with identifier information.
One reason that identifiers are particularly important in Coq is that
Coq has no primitive datatypes; \emph{every} referenced type is an
identifier.
These names can carry a lot of meaning---and that meaning can be
reflected in the names of theorems that refer to them.
This paper describes and evaluates improvements to identifier
encodings in the tactic prediction model.

\paragraph{Categories of Identifiers}

To begin to harness the latent information in identifiers, \toolname
adds three categories of identifiers to the term model of ASTactic.
%
%% Figure~\ref{fig:overview-diagram} shows how \toolname encodes
%% different kinds of identifiers in different ways, to produce a tree
%% which can be encoded into a feature vector.
%% %
%% This paper describes and evaluates improvements to identifier
%% encodings in the tactic prediction model.

\begin{figure}
  \begin{lstlisting}
Definition (@\codesimb{posnat}@) := {(@\codesima{n}@) : nat | n > 0}.

Inductive (@\codesimb{posnatEq}@) : posnat -> posnat -> Prop :=
  | (@\codesimc{posnatEq_intro}@) : ...

Definition (@\codesimb{posnatMult}@)((@\codesima{p1}@) (@\codesima{p2}@) : posnat) : posnat := ...
  \end{lstlisting}
  \caption[Definitions related to the posnat type.]{Definitions
    related to the posnat type, a type of pairs of natural numbers and
    proofs that they are greater than zero. These definitions are
    found in the Foundational Cryptography
    Framework,~\protect\footnotemark{} retrieved as part of the
    Verified Software Toolchain.~\protect\footnotemark{}}
  \label{fig:overview-definitions}
\end{figure}

\addtocounter{footnote}{-1}
\footnotetext{\url{https://github.com/adampetcher/fcf}}
\addtocounter{footnote}{1}
\footnotetext{\url{https://vst.cs.princeton.edu/}}

To understand these identifier categories, consider the definitions in
Figure~\ref{fig:overview-definitions}, from a verified cryptography
library.
\begin{enumerate}
\item The identifier \lstinline{posnat} is a \emph{global definition}
  (highlighted in \codesimb{red}), it can be used by datatypes,
  functions, theorems, or proof scripts, to reference the globally
  defined \lstinline{posnat} datatype.
\item The identifier \lstinline{n} is a \emph{local variable}
  (highlighted in \codesima{orange}), as it can be referenced within
  the local context of this term, but not outside of it.
\item The identifier \lstinline{posnatEq_intro} is a \emph{type constructor}
  (highlighted in \codesimc{yellow}) as it can be referenced in
  datatypes, functions, theorems, and proof scripts to construct a new
  \lstinline{posnatEq} object.
\end{enumerate}

Appendix~\ref{sec:appendix} further details these categories of
identifiers (global definitions, local variables, and constructor
names) and provides intuition through examples for why each category
may be useful to encode in a tactic prediction model.
Appendix~\ref{ssec:enrichment} details the implementation effort
required for enriching a model with these three categories of
identifiers.

\begin{figure}
  \begin{subfigure}[b]{0.45\linewidth}
\begin{lstlisting}[basicstyle=\linespread{0.95}\footnotesize\ttfamily]

Lemma (@\codesimb{posnatMult_comm}@) : forall (@\codesima{p1} \codesima{p2}@),
  (posnatEq (posnatMult p1 p2)
             (posnatMult p2 p1)).
Proof.
  intuition.
  unfold posnatMult.
  destruct p1; destruct p2.
\end{lstlisting}
\caption{A partial proof of \lstinline{posnatMult_comm}.}
\label{fig:proof-script}
\end{subfigure}
  \begin{subfigure}[b]{0.45\linewidth}
  \begin{lstlisting}[basicstyle=\linespread{0.95}\footnotesize\ttfamily]
x : nat
g : x > 0
x0 : nat
g0 : x0 > 0
============================
(@\vb{posnatEq}@) ((@\vc{exist}@) (fun (@\va{n}@) : nat => n > 0)
                   ((@\vb{Nat.mul}@) (@\va{x}@) (@\va{x0}@))
                   ((@\vb{mult_gt_0}@) (@\va{g}@) (@\va{g0}@)))
           (exist (fun n : nat => n > 0)
                   (Nat.mul x0 x)
                   (mult_gt_0 g0 g))
  \end{lstlisting}
  \caption{The proof state at this point in the proof.}
  \end{subfigure}
  \caption{A proof using the definitions in
    Figure~\ref{fig:overview-definitions}, from the same file.}
  \label{fig:overview-proof-state}
\end{figure}

\paragraph{Encodings}

In Figure~\ref{fig:overview-proof-state}, you can see a proof over
these definitions, \lstinline{posnatMult_comm}.
This proof says that multiplication of posnats is commutative, meaning
you can switch the order of the arguments and the result will always
be the same.
Making progress in this proof state requires understanding several
things about the identifiers involved.
\begin{enumerate}
\item The \lstinline{exist} type constructor is a common constructor for
  sigma (existential) types, and there are specialized tactics (like
  \lstinline{exists} and \lstinline{eexists}) for reasoning with those
  objects.
\item The goal type, \lstinline{posnatEq} is related to
  \lstinline{posnat}s and equality.
\item The \lstinline{Nat.mul} function is defined in the Coq's standard
  library, whereas \lstinline{mult_gt_0} is a theorem about it defined
  in the current project.
\end{enumerate}

Understanding these things requires three different approaches:
attaching special signifiers to common identifiers, processing the
individual pieces of identifiers to understand where they connect to
different concepts, and remembering where the definitions being
referenced are defined.

\begin{figure}[ht]
  \includegraphics[width=\textwidth]{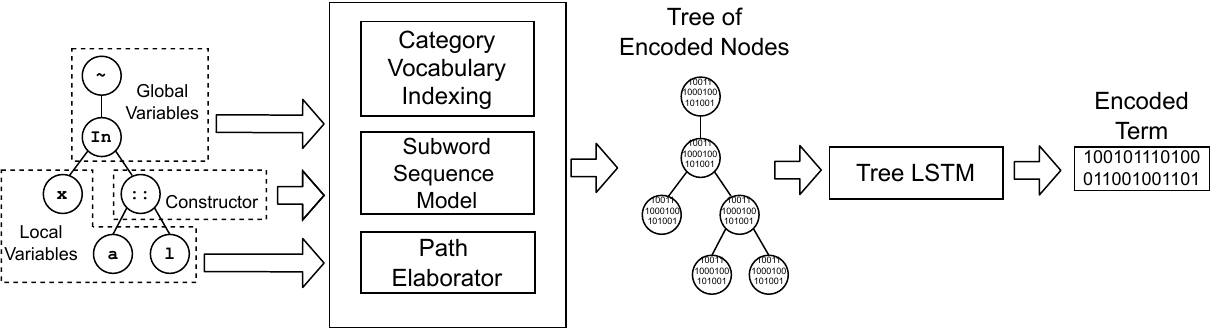}
  \caption{The architecture of \toolname's identifier processing.}
  \label{fig:overview-diagram}
%source: https://drive.google.com/file/d/12sgXdBLl01KRVzufnvOgJDGWapfKUVpq/view?usp=sharing
\end{figure}

The crux of this paper is the enrichment of a proof-synthesis model
for Coq with rich information about
identifiers.
Figure~\ref{fig:overview-diagram} shows an overview of
how identifiers are encoded in \toolname.
To fully take advantage of the richness of these identifiers, our
design employs three key encoding mechanisms:

\begin{enumerate}
\item \emph{Category Vocabulary Indexing}
  (Section~\ref{ssec:common-vocab}), which separately considers
  different kinds of common identifiers in a proof development,
\item \emph{Subword Sequence modeling} (Section~\ref{ssec:bpe}), which draws
  bridges between all identifiers, and
\item \emph{Path Elaboration} (Section~\ref{ssec:paths}), which
  encodes the location where the object referred to by each identifier
  is defined.
\end{enumerate}

Category vocabulary indexing allows us to assign unique labels to common
identifiers in the code.
In this case, that means giving a unique label to the
\lstinline{exist} type constructor, so that we can use knowledge from
previous proofs which used that precise constructor.
Subword sequence modeling allows us to break identifiers up into common
pieces, and process those pieces with a sequence model.
In this case, that means breaking the \lstinline{posnatEq} identifier
into the chunks \lstinline{posnat} and \lstinline{Eq}, so that we can
use knowledge from previous proofs that had identifiers with similar
pieces.
Finally, path elaboration allows us to consider the directories,
files, and modules in which the object referenced by the identifier is
defined.
Here, that means understanding that the multiply identifier refers to
a function defined within \lstinline{Coq.Init.Nat}, but the
\lstinline{mult_gt_0} refers to a lemma defined in the current file.

Armed with the knowledge from these three encoding mechanisms, our
model has everything it needs to learn to complete the proof of
\lstinline{posnatMult_comm}.

\section{Encodings}
\label{sec:encodings}

Identifiers are proxies for semantic information not by accident, but \emph{by design}.
By taking advantage of the information in identifiers, term models can
learn from the design principles the proof engineer has already
followed to make proof developments easier to read, understand, and
build on.
To extract this information from identifiers, \toolname uses three
encoding mechanisms: \textbf{category vocabulary indexing}
(Section~\ref{ssec:common-vocab}), \textbf{subword sequence modeling}
(Section~\ref{ssec:bpe}), and \textbf{path elaboration}
(Section~\ref{ssec:paths}).

\subsection{Category Vocabulary Indexing}
\label{ssec:common-vocab}

In each identifier category (global definitions, local variables, and
type constructors), there are many common identifiers used across
proof developments.
These identifiers are so common that we can learn a significant amount
about how to understand them from their previous uses.
For instance, in the example from Figure
\ref{fig:overview-proof-state}, the \lstinline{exist} type constructor is
part of the standard library, and many proofs in our training data
reason with it.
Even when an identifier is not very common, we can still understand a
lot about it by knowing what category it is in.

To take advantage of these properties of identifiers, we developed
\textbf{category vocabulary indexing}.
This encoding mechanism tags every identifier with the category it comes
from and, if the identifier is commonly used enough, a unique tag for
that particular identifier.
By giving common identifiers a unique tag, we can generalize across
their many appearances, and predict tactics that worked well with them
in the past.
And by marking identifiers with their category, either global
definition, local variable, or type constructor, we can disambiguate
identifiers with the same name from different categories, and learn
useful information about even uncommon identifiers.

Some previous tools for machine-learning-guided proof-synthesis, such
as Proverbot9001~\cite{proverbot} and Tactician~\cite{tactician}, use
vocabulary indexing for common identifiers, but make no category
distinctions.
This is a reasonable approach, because in Coq, the names of global
definitions, local variables, and type constructors share a common
namespace.
However, in \toolname, we decided to distinguish between identifiers
of different categories, in part because manual analysis of
the training data revealed different naming conventions for different
categories.
For example, single-letter identifiers seemed to almost exclusively
represent local variables, with uppercase for types (like
\lstinline{A} in Figure~\ref{fig:option}), and lowercase for terms
(like \lstinline{x} in Figure~\ref{fig:overview-proof-state}); longer
uppercase identifiers generally refer either to sort names (like
\lstinline{Set} or \lstinline{Prop}) or type constructors (like
\lstinline{Some} or \lstinline{None}).
This means that when human provers see an identifier, even if they
haven't seen it before, they often have a sense of what category it
belongs to.

Other previous tools for machine-learning-guided proof-synthesis, such
as ASTactic and TacTok, make category distinctions, but don't index
vocabulary.
We learned early on that the possibility of performance regression due
to uninformative local variables like \lstinline{x} had concerned the
ASTactic authors, and contributed to their decision not to encode
identifiers.~\footnote{\url{https://github.com/princeton-vl/CoqGym/discussions/60}}
However, upon closer inspection of the data we determined that even
when a particular name does not always refer to the same definition,
common names can carry information of their own.
For instance, variables named \lstinline{hd} and \lstinline{tl}
consistently refer to the head and tail of a list.
These names, too, can benefit from a unique tag which generalizes
across their usages.
Our manual inspection determined that this can often hold even for
single-character variable names.

\paragraph{Implementation}
To decide which identifiers are common enough to be indexed, we use
our training data set to create a fixed identifier vocabulary.
That is, we count the occurrences of each identifier, and include in
our vocabulary those whose count is above an experimentally chosen,
fixed threshold (see Section~\ref{ssec:hyperparameters} for an evaluation of
different thresholds).
Using separate vocabularies for each category of identifier allows us
to use different thresholds across different categories; since
type constructors are less common overall than local variables, they might
require having a lower threshold for being included in the vocabulary.

%% Practically, this meant two things: (1) we encoded the category of identifier
%% explicitly in the AST, and (2) we maintained and used separate vocabularies
%% for each of the three categories of identifiers.

%% This choice to distinguish between identifier from different
%% categories means that we store and encode not just one, but three
%% different vocabularies, and three \lstinline{unknown-ident}
%% tokens---one for each category.
%% %
%% It also allowed us to experiment with different thresholds for unknown
%% tokens across different categories.
%% %
%% For out-of-vocabulary identifiers seen at inference time, our model then
%% draws from the distribution of identifiers seen below the threshold
%% \emph{for that particular category} in training data, effectively
%% making use of the fact that the identifier falls into that category.

\subsection{Subword Sequence Modeling}
\label{ssec:bpe}

Identifier information can be useful not just for learning about
individual datatypes, theorems, and functions, but also for drawing
bridges between them.
Developers often organize development using parts of names to group
theorems and functions which refer to common definitions.
It turns out these naming conventions can be useful to a model, too.

Many variable names are not simply single unique words, but are made
up of multiple parts.
These parts could be multiple english words in camel case, such as the
case in something like \verb|firstItemInList| broken into ``first'',
``item'', ``in'', and ``list''. Or they could be components of a
single word that carry individual meaning, like
\verb|prelocalizations| broken into ``pre'' ``local'' ``ization''
``s''.
By breaking these identifiers into pieces, \toolname can learn the
meaning of shared pieces and generalize across identifiers.

In the example from Section~\ref{sec:overview}, \toolname breaks
\lstinline{posnatMult} into \lstinline{[pos, nat, Mult]}; with a
different subword vocabulary, from a different set of variable
occurrences in the training data, it might produce \lstinline{[posnat, Mult]}.
%% %
%% With this subword vocabulary, even though ``posnat'' is not in the
%% vocabulary, all instances of ``posnat'' will be broken down into the
%% same two tokens ``pos'' and ``nat'', so the sequence model will be
%% able to link them.
%
These tokens are processed with a sequence model, so that the
identifier's ultimate feature vector reflects the fact that the
identifier relates to the ``posnat'' type, and that it primarily
relates to the multiplication operation.

To get a sense for this, let us consider another example.
The Coq standard library includes operations
about the real numbers \lstinline{R}, like addition:

\begin{lstlisting}
  (@\codesimb{Rplus}@) : R $\rightarrow$ R $\rightarrow$ R.
\end{lstlisting}
The library contains proofs of theorems about \lstinline{Rplus},
like this proof (highlighting just one \lstinline{Rplus} for presentation):

\begin{minipage}{\linewidth}
\begin{lstlisting}
  Lemma Rplus_eq_compat_l : $\forall$ (r r1 r2 : R),
      r1 = r2 $\rightarrow$ (@\codesimb{Rplus}@) r r1 = Rplus r r2.
  Proof.
    intros r r1 r2.
    apply f_equal.
  Qed.
\end{lstlisting}
\end{minipage}
which proves the theorem that right addition preserves equality.

Suppose we wish to prove the analogous theorem about the natural
numbers \lstinline{nat}, using the addition function \lstinline{plus}
defined over \lstinline{nat}.
We can do this the same way:

\begin{minipage}{\linewidth}
\begin{lstlisting}
  Lemma plus_eq_compat_l : $\forall$ (n n1 n2 : nat),
      n1 = n2 $\rightarrow$ (@\codesimb{plus}@) n n1 = plus n n2.
  Proof.
    intros n n1 n2.
    apply f_equal.
  Qed.
\end{lstlisting}
\end{minipage}
simply renaming the local variables for style (though the original
proof with \lstinline{r}, \lstinline{r1}, and \lstinline{r2} also
works with no changes).

The fact that \lstinline{Rplus} and \lstinline{plus} are related is
explicit in the identifier names: \lstinline{Rplus} behaves like
\lstinline{plus} over \lstinline{R}.
A model that can draw connections between \lstinline{plus} and
\lstinline{Rplus} can in some cases reuse proofs about one to derive
analogous proofs about the other.

The key here is subword sequence modeling which excels at
drawing connections between related words~\cite{bpe-1, bpe-2}.
Subword sequence modeling allows us to break the identifier
\lstinline{Rplus} into the chunks \lstinline{R} and \lstinline{plus},
and index them separately, connecting them to the identifier
\lstinline{plus}.
By drawing these connections, we expect that a model can suggest
\lstinline{intros} and \lstinline{f_equal} in the body of
\lstinline{plus_eq_compat_l}, by connecting the hypothesis
\lstinline{plus n n1 = plus n n2} to the hypothesis
\lstinline{Rplus n n1 = Rplus n n2}.
With subword sequence modeling, the model can learn all of this with
no need for semantic information about what each of the reals and
naturals represent, or how their addition functions are related.

In \toolname, identifiers are broken into subwords using a byte-pair
encoding algorithm (BPE)~\cite{bpe-1, bpe-2}, an algorithm that has
seen success in code completion models for program
synthesis~\cite{big-code, fast-code}.
The algorithm uses the training corpus to make a list of common
subwords by starting with a vocabulary of single characters, and
iteratively merging common pairs.
Then, each identifier is tokenized by greedily consuming the longest
matching vocabulary element.

Passport incorporates these tokens as embeddings in a syntax model.
Program syntax can generally be modeled in two ways.
The simplest way is to model it as an unstructured sequence of words
(or more generally, tokens).
The alternative is to parse the syntax into a tree, and use a tree
based model to process it.
One of the advantages of the former is that you can tokenize strings
in a number of different ways, including with multiple tokens per
identifier (sub-word tokenization).
However, \toolname builds on a parsed-tree based model, so there is no
existing string tokenizer which could be used for subword
tokenization.
Instead, we embed a sequence model \emph{within the leaves} of the
tree-based syntax model.
This means that our subword sequence model only learns how to combine
parts of an identifier into a fixed embedding for the identifier, and
doesn't need to learn about other parts of program syntax.

With our category vocabulary indexing, we used separate vocabularies for
identifiers of different categories.
However, proof developments sometimes demonstrate connections between
identifiers from different categories.
These connections are lost in using separate vocabularies, so subword
encoding is used to maintain these connections.
In \toolname, we use a single subword vocabulary, derived from the
global variable corpus, to encode identifiers from all categories.

\paragraph{Implementation}
There are several subtleties to the implementation of our subword
tokenization algorithm, and the byte-pair encoding which generates
its vocabulary.
Sometimes there were several possible ways to implement the approach;
in general, we made our choices based on the performance of the
resulting model on our benchmarks.

As indicated by the name, byte-pair tokenization often starts with a
vocabulary of bytes, not characters, to allow a reasonable base
vocabulary size when working with unicode.
However this has the downside of sometimes indicating that two
identifiers are similar because they share bytes within a unicode
character, even if no characters are in common.
In our implementation, we use characters as our base vocabulary.
To keep our base vocabulary of a reasonable size, we only include
those characters which are present in the training corpus.
Since Coq programmers generally only use a small subset of possible
unicode characters, this works well.
However, there are in rare cases unicode characters present in the
test data which are not present in the training data.
To address this, our subword tokenizer drops characters which are not
present at all in the vocabulary; this behavior can be changed with a
flag to instead produce a special \lstinline{<unknown>} element.

Many different neural architectures have been used to process
sequences of tokens.
For language modeling, the most effective models are often those with
attention and forgetfulness mechanisms, to capture the long-range
dependencies present in text.
However, the identifiers we work with are generally short, often only
a few subwords long, so we instead use the simplest sequence model, a
Recurrent Neural Network, without any attention mechanism.

As with any sequence-based model, there is a question of how to cap the
size of sequences so that their lengths can be normalized.
With \toolname, we found that capping at four tokens per identifier
during training, but eight tokens per identifier when synthesizing
proofs, is most effective on our evaluation suite.

\subsection{Path Elaboration}
\label{ssec:paths}

The final encoding mechanism in \toolname is path elaboration: the
encoding of fully-qualified paths of different identifiers.
By paying attention to the fully-qualified paths of different
identifiers, \toolname can take advantage of any grouping of
identifiers into common modules and files already used by Coq
developers to organize development.
\toolname can also capitalize on proof development styles that dispatch proofs
for entire classes of related theorems using powerful tactics---a proof
development style recommended by, for example, the popular Coq textbook
Certified Programming with Dependent Types~\cite{cpdt}.

To gain some intuition for what this means in action,
consider this proof of a theorem from the Coq standard library:

\begin{minipage}{\linewidth}
\begin{lstlisting}
  Theorem not_in_cons A (x a : A) (l : list A):
    ~ In x (a::l) $\leftrightarrow$ x<>a $\land$ ~ In x l.
  Proof.
    simpl. intuition.
  Qed.
\end{lstlisting}
\end{minipage}
%% The theorem states that for any type \lstinline{A},
%% for any two elements \lstinline{x} and \lstinline{a} of type \lstinline{A},
%% and for any list \lstinline{l} of elements of type \lstinline{A},
%% these two statements imply one another:

%% \begin{enumerate}
%% \item \lstinline{x} is not in the list formed by consing \lstinline{a} onto \lstinline{l}, and
%% \item \lstinline{x} is not \lstinline{a}, and \lstinline{x} is not in \lstinline{l}.
%% \end{enumerate}

The proof of \lstinline{not_in_cons} goes through by just two tactics: \lstinline{simpl}
and \lstinline{intuition}.
The \lstinline{simpl} tactic simplifies the initial goal (no assumptions, with the theorem type as the sole proof obligation) to
make it easier to reason about, producing this proof state:

\begin{minipage}{\linewidth}
\begin{lstlisting}
  A : Type
  x, a : A
  l : list A
  ______________________________________(1/1)
  ~ (a = x $\lor$ In x l) $\leftrightarrow$ x <> a $\land$ ~ In x l
\end{lstlisting}
\end{minipage}
In this case, the \lstinline{simpl} tactic has unfolded the \lstinline{In x (a::l)} on the left side of the identifier into \lstinline{(a = x $\lor$ In x l)}.

But the resulting goal is still a bit complex because it chains together a
number of logical connectives: if and only if ($\leftrightarrow$), negation (\lstinline{~}), inequality (\lstinline{<>}), conjunction ($\land$), and disjunction ($\lor$).
So the \lstinline{intuition} tactic breaks down logical connectives
into simpler subgoals, and dispatches each subgoal automatically.

Taking a step back, it is natural to wonder how the proof engineer
could have known to use the \lstinline{intuition} tactic to dispatch
the remaining goals.
Intuitively, it made sense to use \lstinline{intuition} here because
the goal consisted of simple statements linked by logical connectives,
which \lstinline{intuition} excels at.
It turns out that the fact that these operators are logical
connectives is explicit in the paths of the identifiers in the
goal---they all reside in the \lstinline{Coq.Init.Logic} module---so
we can pass it on to \toolname by encoding paths.

We can see this by expanding the paths of the identifiers in the theorem
statement of \lstinline{not_in_cons} (Figure~\ref{fig:notincons}).
All of the operators in \lstinline{not_in_cons} are syntactic sugar
for identifiers, which themselves refer to types defined inductively
in Coq.
For example, conjunction ($\land$) refers to the inductive type
\lstinline{and} in the path \lstinline{Coq.Init.Logic}.
% (using the
%same conventions as in Section~\ref{sec:introduction} to denote
%different categories of identifiers):
%
%\begin{lstlisting}
%  Inductive (@\codesimb{and}@) ((@\codesima{A B}@) : Prop) : Prop :=
%  | (@\codesimc{conj}@) : A $\rightarrow$ B $\rightarrow$ and A B
%\end{lstlisting}
%This type has a single constructor \lstinline{conj}, which takes a term of type
%\lstinline{A} and a term of type \lstinline{B}, and from that constructs
%a term of type \lstinline{and A B}.
%In other words, if \lstinline{A} is true and \lstinline{B} is true,
%then \lstinline{and A B} is true.
Internally, Coq stores the elaborated theorem with all of these identifiers
(like \lstinline{and}) and their fully-qualified paths
(like \lstinline{Coq.Init.Logic}) explicit.
Inspecting the elaborated version of \lstinline{not_in_cons} shows
that the fact that these are logical connectives requires no semantic
understanding to deduce---it is explicit in the grouping of
identifiers in the \lstinline{Logic} module.

\begin{figure}
  \begin{lstlisting}
  (@\codesimb{not_in_cons}@)
  : (@\ltacforall@) ((@\codesima{A}@) : Type) ((@\codesima{x a}@) : A) ((@\codesima{l}@) : list A),
      Coq.Init.Logic.(@\codesimb{iff}@)
        (Coq.Init.Logic.(@\codesimb{not}@)
          ((@\codesimb{In}@) A x ((@\codesimc{cons}@) A a l)))
        (Coq.Init.Logic.(@\codesimb{and}@)
          (Coq.Init.Logic.not
            (Coq.Init.Logic.(@\codesimb{eq}@) A x a))
          (Coq.Init.Logic.not (In A x l))).
  \end{lstlisting}
  \cprotect\caption{The theorem statement \verb|not_in_cons|,
    elaborated with paths. Highlighted using the same conventions as
    in Figure~\ref{fig:overview-definitions}, with other paths omitted for
    brevity.}
  \label{fig:notincons}
\end{figure}
%% The decision to include fully-qualified paths was inspired by
%% exploration of proofs in the data that used high-level tactics to
%% dispatch similar goals.
%% %
%% When walking through proofs that used \lstinline{intuition}, for
%% example, we found ourselves not looking at particular subterms like
%% $\land$ and $\lor$, but rather getting a sense that the entire goal
%% was a collection of logical statements that would best be handled by a
%% high-level logic tactic like \lstinline{intuition}.

We determined that a simple way to pass this intuition on to \toolname
was to encode each of the file and module names inside
of fully-qualified paths, taking advantage of the organization of
large proof developments to infer tactics used to dispatch related goals.

\paragraph{Implementation}
To implement this, we created a dedicated vocabulary and corresponding unknown
for file and module names inside of fully-qualified paths, much like we did for each
category of identifier.
We then used this vocabulary for encoding paths.

As with identifiers, Coq includes fully-qualified paths inside of the ASTs by default,
but TacTok and ASTactic had erased those paths from the AST.
For example, in Figure~\ref{fig:constructor}, the fully-qualified
path \lstinline{Coq.Init.Datatypes} of the \lstinline{option} inductive type
shows up in the AST as a \lstinline{directory_path} node,
with data \lstinline{[Datatypes; Init; Coq]}.

Elaborating paths was thus similar to adding each of the categories of identifiers:
First, we modified the post-processing code to avoid erasing paths.
Then, we built a separate vocabulary for common files or modules that paths
consisted of, like \lstinline{Datatypes}, \lstinline{Init},
and \lstinline{Coq} in Figure~\ref{fig:constructor}.
We then encoded each file or module along the path separately,
mapping to a dedicated unknown token for files or modules in paths
that occurred less frequently than the chosen threshold.

\section{Evaluation}
\label{sec:evaluation}

We evaluated \toolname's ability to successfully prove theorems using
the CoqGym benchmark~\cite{Yang19}, following the evaluation
methodology used by several recent papers~\cite{Yang19, First20, First22icse}.

%% We found that the combined proving power of the \toolname-enhanced
%% models exceeds that of the original models by 38\%.
%% %
%% Combining both the \toolname-enhanced and unenhanced models does even
%% better, outperforming the combined unenhanced models by 45\%
%% (Section~\ref{ssec:main-experiment}).

%% Next, we measured the impact of encoding each category of identifier
%% separately, without the others (Section~\ref{ssec:eval-categories}).
%% %
%% Then, we perform an ablation study on several of our encoding
%% mechanisms, including their impact on models that encode only one
%% category of identifier
%% (Sections~\ref{ssec:subwords-eval}~and~\ref{ssec:paths-eval}).
%% %
%% We find that each identifier category and encoding mechanism
%% individually increases the number of theorems our model can prove, but
%% the effect of including them all is greater than their combined
%% individual effects.

%% We also studied how the problem of nondeterminism in model
%% training~\cite{pham-repro, google-ai-repro, fairness-repro} affects
%% the performance and reliability of machine-learning-based proof-search
%% tools.
%% %
%% We found that even with the same model and training data, proof search
%% success rate can vary by 0.4\%, or 38 proofs in our dataset
%% (Section~\ref{ssec:nondeterminism}).

%% Finally, we measured the impact of changing various hyperparameters in
%% our model (Section~\ref{ssec:hyperparameters}).

A summary of the results is as follows:

\begin{itemize}
\item \textbf{\toolname improves proving power.}  By comparing to
  previous tools---ASTactic and the two models, Tac and Tok, that make
  up TacTok---we measured additional proving power provided by
  \toolname's model of identifiers.  The combined proving power of the
  \toolname-enhanced models exceeds that of the original models by
  38\%, and combining both the \toolname-enhanced and unenhanced
  models outperforms the combined unenhanced models by 45\%
  (Section~\ref{ssec:main-experiment}).
\item \textbf{Identifiers improve performance.} All three categories
  of identifiers improve performance, in aggregate proving 64\% more
  theorems than the individual unenhanced model
  (Section~\ref{ssec:eval-categories}).
\item \textbf{All three encoding mechanisms improve performance.} The
  models enhanced with all three categories of identifiers perform
  better with each of the three \toolname encoding mechanisms
  (Sections~\ref{ssec:subwords-eval}~and~\ref{ssec:paths-eval}).
\item \textbf{Our results are meaningful beyond variance introduced by
  nondeterminism.}  Proof search success rate varies by 0.4\% for
  individual models, and combining many varying runs can improve
  results by 22\% (Section~\ref{ssec:nondeterminism}).
\item \textbf{Hyperparameter choices impact performance.}  We choose
  our hyperparameters experimentally based on these results
  (Section~\ref{ssec:hyperparameters}).
\end{itemize}

\subsection{Experimental Setup}
\label{sec:setup}

\paragraph{Benchmark}
The CoqGym benchmark includes 124 open-source Coq projects, split into
three sets.
For our evaluation, we trained on 97 projects (containing a total of 57,719 theorems) and synthesized proofs for 26 projects (containing a total of 10,782 theorems).
We exclude one project, coq-library-undecidability, from our evaluation because
TacTok's evaluation~\cite{First20} was unable to reproduce prior results for
ASTactic's performance~\cite{Yang19} on that project due to internal Coq
errors when processing the proof scripts.

Projects in the CoqGym benchmark are a mixture of mathematical
formalizations, proven correct programs, and Coq automation libraries.
They include several compilers of varying sizes (such as CompCert~\cite{compcert}),
distributed systems (such as Verdi~\cite{verdi}),
formalizations of set theory, and more.
Some of the projects in CoqGym (such as the automation libraries)
do not contain any proofs, but we included them for completeness.

\paragraph{Machines}

We ran this paper's experiments using two clusters: a GPU cluster
for training and a CPU cluster for synthesizing proofs.

Each node in the GPU cluster has between two and eight NVIDIA GPU
cards.
There are four nodes with two NVIDIA Tesla V100 GPUs, and thirty-three
nodes with eight NVIDIA RTX 2080ti GPUs.
The nodes in the GPU cluster all run on a shared ZFS file system, run
CentOS Linux, and use Slurm for job scheduling and resource
management.

Each node in the CPU cluster has between 24 and 36 cores, with 4
hyperthreads per core.
There are:
\begin{itemize}
\item 1 head node with 24 cores of Xeon E5-2680 v4 @ 2.40GHz, 128GB
  RAM and 200GB local SSD disk.
\item 50 compute nodes with 28 cores of Xeon E5-2680 v4 @ 2.40GHz,
  128GB RAM and 200GB local SSD disk.
\item 50 compute nodes with 28 cores of Xeon Gold 6240 CPU @ 2.60GHz,
  192GB RAM and 240GB local SSD disk.
\item 5 compute nodes with 56 cores of Xeon E5-2680 v4 @ 2.40GHz,
  264GB RAM and 30TB local disk.
\end{itemize}
The nodes in the CPU cluster also all run on a shared ZFS file system,
run CentOS Linux, and use Slurm for job scheduling and resource
management.

\paragraph{Experimental Parameters}

\toolname attempts to synthesize each proof for a preset amount of
time, timing out if it fails to to reach \lstinline{Qed} in that time.
Our evaluation used 10 minutes for this timeout, following the choice
made by ASTactic~\cite{Yang19} and TacTok~\cite{First20}.
Our experiments use 200 as the default category vocabulary threshold
(recall Section~\ref{ssec:common-vocab}) and 4,096 as the default
byte-pair merge threshold (recall Section~\ref{ssec:bpe}).
We use 128 as the default vector dimension for term, grammar, and
terminal/non-terminal symbol embeddings, as well as the
dimension of the LSTM controller. For all other parameters, we follow
those used by ASTactic~\cite{Yang19} and TacTok~\cite{First20}.

\subsection{\toolname's Effect on Proof-Synthesis Tools}
\label{ssec:main-experiment}

In this section, we show that the addition of our identifier
information improves the end-to-end performance of proof search tools.
Since \toolname is implemented in the ASTactic/TacTok framework, we
were able to evaluate our changes against three base models:
An ASTactic-like\footnote{We were not able to replicate the original
results of the ASTactic model~\cite{Yang19}, so for our evaluations we
trained this model with the same embedding vector dimensions as our
own models. For this reason we are using the term ASTactic-like when
we describe our results.} model, Tac, and Tok.
ASTactic was developed as part of the CoqGym project~\cite{Yang19},
and uses only proof contexts as input to their prediction model.
By contrast, the Tac and Tok models (developed as part of the TacTok
project~\cite{First20}) additionally model the proof script up to the
current point, with the Tac model encoding the tactics in the proof
script, and the Tok model encoding all the tokens except punctuation
in the proof script.

\begin{figure}[t]
  \includegraphics[width=\columnwidth]{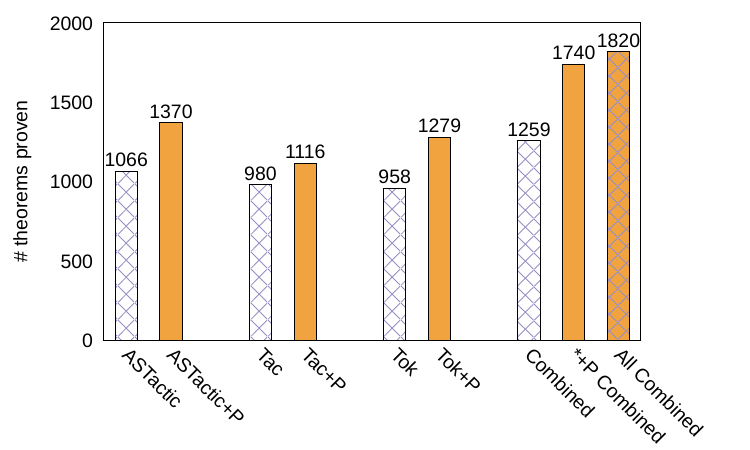}
  \caption{The effect of adding all of \toolname's three encodings for three
    identifier types to several proof-synthesis models. The purple
    crosshatch bars represent baseline models based on ASTactic, Tok,
    and Tac. The orange bars represent our new contributions. The
    rightmost crosshatch bar, labeled ``Combined'', is the number of
    theorems successfully proven by \emph{at least one} of the
    baseline models. The orange bar next to that, labeled ``*+P
    Combined'', is the number of theorems successfully proven by
    \emph{at least one} of the \toolname-enhanced models. Finally, the
    orange \emph{and} crosshatched bar on the far right is the number
    of theorems proven by at least one of all the presented models.}
  \label{fig:main-eval-results}
\end{figure}

Figure~\ref{fig:main-eval-results} shows the results of adding
identifier information to all three of these models.
Adding identifiers to each of the three models significantly improves their
ability to prove theorems.
Adding identifier information improves our ASTactic-like model by 29\%
(304 additional theorems proved), Tac by 14\% (136 additional theorems proved), and Tok by
33\% (318 additional theorems proved).

Following TacTok's~\cite{First20} and Diva's~\cite{First22icse} evaluations, we also
explore how the differences in theorems proven by multiple models
lead to more theorems proven overall, and how adding identifier
information increases that improvement.
When we union the proofs synthesized by all our \toolname-enhanced models,
and compare that set to the union of the proofs synthesized by the base models,
we find an improvement of 38\%.
Comparing the union of theorems proven by all the models to the union of theorems proven
by the three base models, we find an improvement of 45\%.

Next, we examine the complexity of the proofs \toolname generated.
Using human-written proof-script length as a rough proxy for
complexity, we note that \toolname successfully synthesized proof
scripts for 353 theorems for which the human-written proof scripts
were at least 5 tactics long. For 65 of those theorems, the
human-written proof scripts were at least 10 tactics long.
This observation
suggests that \toolname is able to synthesize a significant number of
nontrivial proofs. For 283 theorems, \toolname was able to synthesize
proof scripts that were shorter than the human-written ones. In one
particular case, the human-written script was 139 tactics long, while
\toolname's script was only 2 tactics long.
% (using the tauto tactic).

Examining the time it takes \toolname to synthesize a proof script, the
successfully generated proof scripts took between 0.08 and 86.6~seconds to
generate, with the mean of 2.9~seconds.

%% Adding identifiers to the ASTactic model does not appear to improve
%% its ability to prove.
%% %
%% However, we can see that adding identifiers to the Tok model takes it
%% from a model that proofs a complementary but smaller set of proofs
%% than ASTactic to one that proves more proofs.
%% %
%% It's possible that this is because identifier information is primarily
%% useful in conjunction with information about the proof so far.
%% %
%% It's important to note however that the number of proofs solved for
%% ASTactic is taken from the ASTactic paper; we were not able to
%% independently reproduce this result.
%% %
%% Therefore, it's possible that the decrease in proofs solved between
%% ASTactic and ASTactic with identifiers is due primarily to differences
%% in our evaluation setups; we noticed when working with the ASTactic
%% codebase several instances of non-determinism in the training and
%% testing procedures, several of which were machine and file-system
%% ordering dependent.

\subsection{Identifier Categories}
\label{ssec:eval-categories}
\toolname models several categories of identifiers.
While the experiment in Section~\ref{ssec:main-experiment} showed that
modeling identifiers from these categories are effective together, we
also wanted to show the utility of the identifier categories
individually.

\begin{figure}
  \begin{subfigure}[t]{.45\linewidth}
    \includegraphics[width=\linewidth]{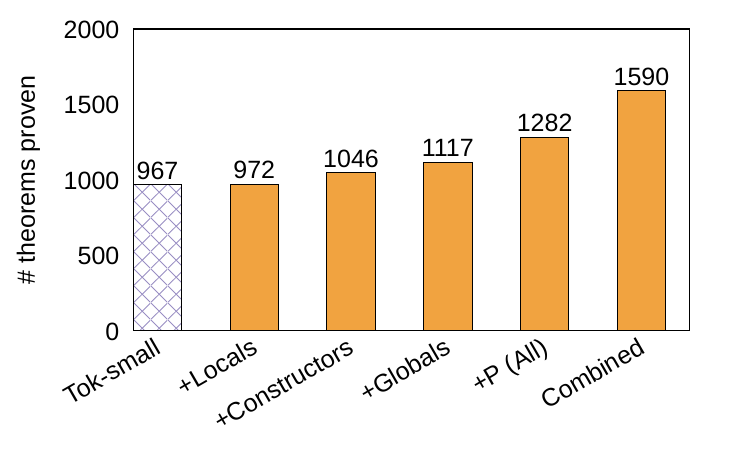}
    \caption{The impact of category vocabulary indexing on three
      identifier categories (without subwords or paths): local
      variables, type constructors, and global definitions.}
    \label{fig:individual-techs}
  \end{subfigure}
  \hspace{1em}%
  \begin{subfigure}[t]{.45\linewidth}
    \includegraphics[width=\linewidth]{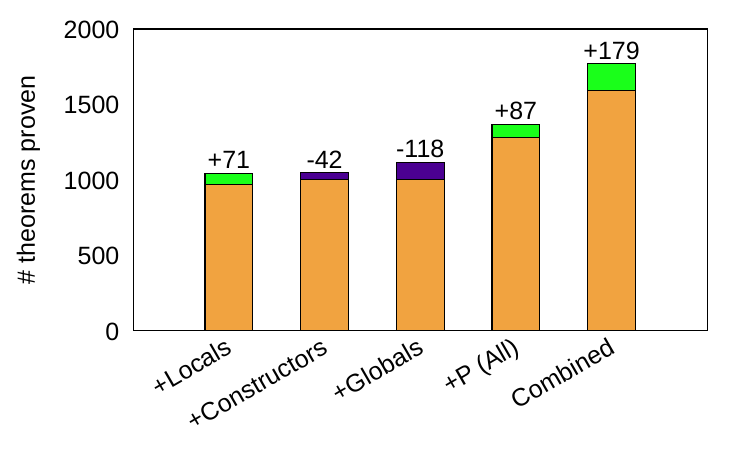}
    \caption{The impact of subword encoding on each of the categories
      of identifiers (with category vocabulary indexing but without
      paths).}
    \label{fig:bpe}
  \end{subfigure}
  \begin{subfigure}[t]{.45\linewidth}
    \includegraphics[width=\columnwidth]{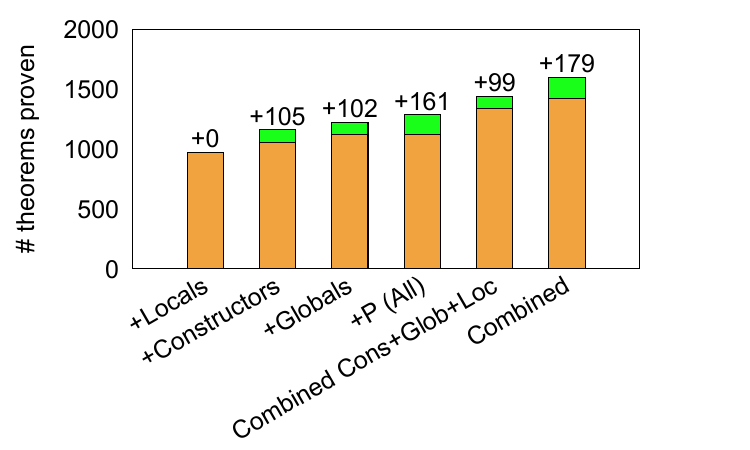}
    \caption{The impact of fully-qualified path encoding of
      type constructors and global definitions (with category vocabulary
      indexing but without subwords).}
    \label{fig:path-results}
  \end{subfigure}
  \caption{}
  \label{fig:category-results}
\end{figure}

Figure~\ref{fig:individual-techs} shows the individual results of just
adding local variables, type constructors, and global definitions.
For consistency, this experiment compares to a Tok-like model with smaller
embedding sizes, as \toolname uses that model to add identifier information
to.

Each of the identifier types added individually
increases the number of theorems proven, though the increase from
local variables alone is marginal.
Adding type constructors alone proves 8\% more theorems than the
baseline, adding global definitions alone proves 16\% more theorems,
and adding local variables alone proves 0.5\% more theorems.

However, no identifier category added individually is close to the impact of
adding all three.
Adding all three identifier types, without subword information,
proves 33\% more theorems.

Finally, though none of the models with individual identifier types
prove as many theorems as the one with all of them together, some
of these individual identifier models prove theorems that the all-identifiers
model does not.
The union of the theorems proven by the individual identifier models
and the all-identifiers model contains 64\% more theorems than the
baseline model.

These experiments show that each identifier category is useful for
producing a more effective proof-synthesis model, and that the
identifier categories help with a diverse set of theorems, so
combining the results of adding different subsets of identifiers helps
further.

\subsection{Subwords}
\label{ssec:subwords-eval}
%% Each model enrichment explored in Section~\ref{sec:idents}
%% encodes a new type of identifier, using a fixed index for the most
%% common identifiers of that type.
%% %
%% Section~\ref{ssec:bpe} described how we augmented this
%% information using subword encodings for each identifier.

Figure~\ref{fig:bpe} shows the impact of adding subword encodings
to our identifier models (Section~\ref{ssec:bpe}).
Adding the subword encoding does not benefit all types of
identifiers individually.
In fact, it makes two (type constructors and global definitions) out
of the three identifier categories perform worse than when those
identifiers are used individually, possibly due to overfitting.
%
% This is not completely surprising, as adding more information to a
% model can cause it to overfit.

However, when subwords are added to the full model with all the
identifier categories, they improve results by 7\%.
This improvement is greater than what the cumulative impact of adding
subwords to individual identifier models, suggesting that subwords
particularly help with making connections between multiple identifier
types.
In fact, even though subword sequence modeling doesn't help global
definitions alone, when global definitions are combined with the other
identifier types, removing subword encoding significantly hurts
results.

The most likely explanation for these results is that for subwords to be
effective, a sufficiently large number of identifiers is necessary to
encounter a non-trivial number of repeated subwords, allowed for learning
semantics of those subwords. Adding subwords to only a single type of
identifier likely does not meet that threshold, but using all identifiers
leads to a significant improvement in the model's proving power.

%% As you can see, adding subword encoding using BPE benefits all types
%% of identifiers.
%% %
%% However, the impact of subword encoding differs greatly across
%% identifier types.
%% %
%% In fact, subword encoding benefits global variables more than twice as
%% much as it does local variables.
%% %
%% Part of the reason for this is that global identifiers often have
%% longer, more informative names.
%% %
%% But part of it is also that, due to the shared scope of all global
%% variables, they're more likely to refer to each other in their names.

\subsection{Paths}
\label{ssec:paths-eval}
Figure~\ref{fig:path-results} shows the impact of removing path
elaboration (Section~\ref{ssec:paths}) from various identifier types
in the \toolname model.
Since local variables do not have paths, there is no impact of
removing path elaboration.
Subwords were not included in this experiment, as we wanted to isolate
the impact of paths.

Path elaboration benefits both type constructors and global
definitions: increasing proofs solved for type constructors alone by
10\% and increasing proofs solved for global definitions alone by 9\%.
When the proofs solved for these categories alone are unioned with the
proofs solved with local variables alone (for which the paths improvement is
0\%), adding path elaboration improves the result by 7\%.
However, when we add path elaboration to \toolname with \textit{all
  three} identifier categories, it increases the number of proofs
solved by 12.6\%.

%% to the \textit{combination} of the the The change
%% for the model including all three identifiers was 14\%. We also
%% examined the combination of all theorems proved by the three models
%% using each identifier individually, as well as the combination of
%% these three models and the model that uses all identifiers. Paths
%% improved the former by 7\%, and the latter by 12.6\%. These
%% percentages are lowered due to the total lack of improvement for the
%% locals category.
These results indicate that the impact of adding path elaboration to a
model that implements local variables, type constructors, and global
definitions is greater than the combined effect on individual models.
Similarly to the subword experiment above, these results suggest that
encoding fully-qualified paths helps connect identifiers across
categories;
learning about how type constructors from a particular module behave
helps in dealing with global definitions from that module, and visa
versa.
However, unlike the subword experiment, paths seem to benefit all
identifiers for which they are implemented individually as well as in
combination.

\subsection{Non-Deterministic Model Variance}
\label{ssec:nondeterminism}

During the course of evaluating our project, we found that models
trained in the ASTactic framework had significant variance in
their proof-synthesis success rate, even when the model code and training data were
identical.
While part of this variance could be attributed to different hardware
and other hard-to-control factors (see Section~\ref{sec:discussion}),
even when controlling for all those factors, there was still variance.
After months of investigation, we found that the cause was
non-determinism at the hardware and framework level, some of it
undocumented~\cite{pytorch-repro-bug, cub-repro-bug}.

Non-determinism in model training is not specific to proof search, and
has in fact been documented in the ML community at
large~\cite{pham-repro, google-ai-repro, fairness-repro}.
However, it is not immediately obvious how these effects would impact
proof search, since they are usually measured as inaccuracy in the top
prediction of a model, while proof search tools generally use multiple
model predictions, smoothing out some inaccuracy.

To measure the impact of non-deterministic training variance on proof
search, we trained our model with identifiers added to Tok 20 times.
On average, the models proved 11.9\% (1,279 theorems), with
the maximum proving 12.0\% (1,294 theorems) and the minimum proving
11.6\% (1,256 theorems). The 0.4\% spread (38 theorems) shows that
training the same model can lead to small differences in overall success rates.
Our result for adding local variables alone (with no other
identifiers) and without subword encoding is within this variance
range.
However, the impact of local variables is better captured with the
addition of subwords and together with other identifiers, which yields
results significantly outside of this range.

Interestingly, the union of the theorems proven by the 20 models is
14.5\% (1,564 theorems), an improvement of 22\% over the average. This
demonstrates that the scale of the differences in \emph{which}
theorems models can prove as a result of non-deterministic training
variance is much larger than the scale of the differences in \emph{how
many} they prove. Thus, the variance from training non-determinism
serves as a dimension for model diversity, which can be used to
improve proof synthesis, similarly to the approach taken by
Diva~\cite{First22icse}.

\subsection{Hyperparameters}
\label{ssec:hyperparameters}

\begin{figure}
  \begin{subfigure}[b]{.45\linewidth}
    \includegraphics[width=\linewidth]{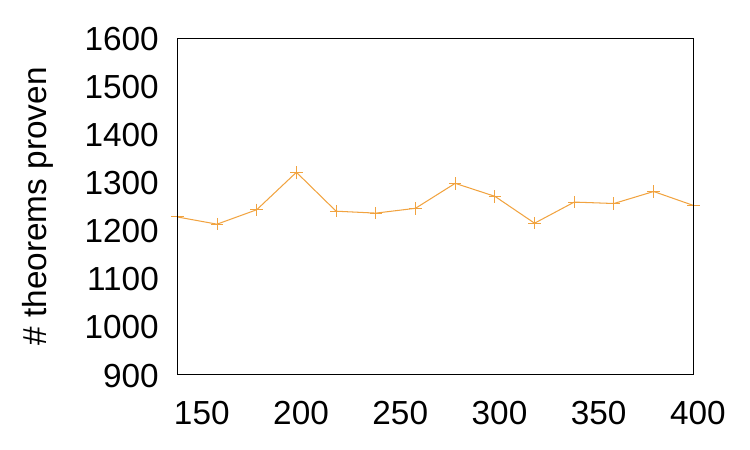}
    \caption{Global definitions}
  \end{subfigure}
  \begin{subfigure}[b]{.45\linewidth}
    \includegraphics[width=\linewidth]{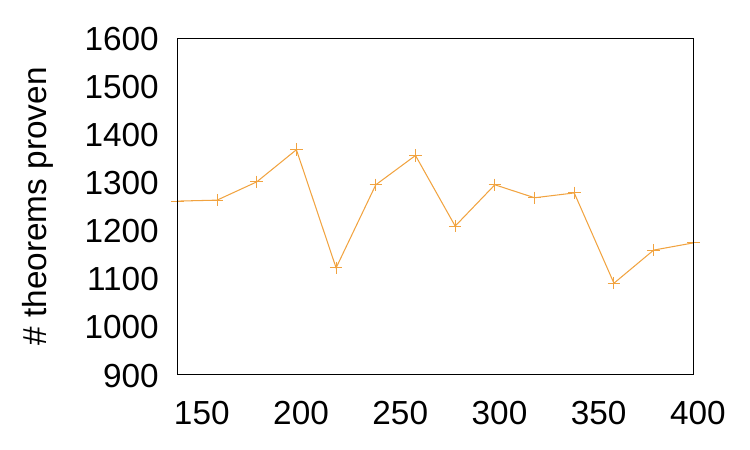}
    \caption{Local variables}
  \end{subfigure}

  \begin{subfigure}[b]{.45\linewidth}
    \includegraphics[width=\linewidth]{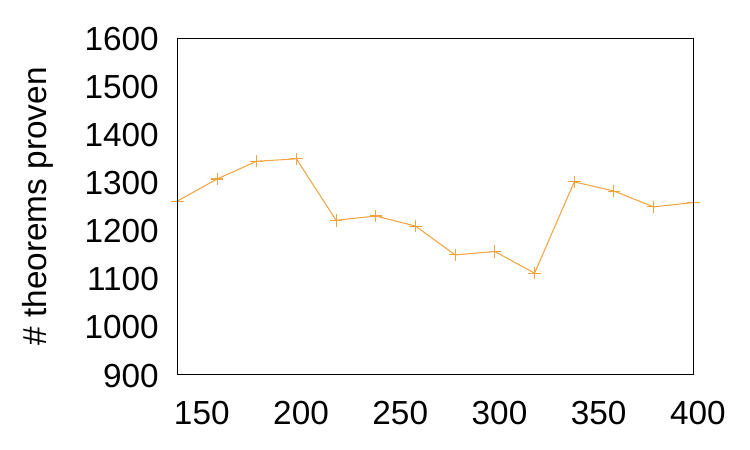}
    \caption{Type constructors}
  \end{subfigure}
  \begin{subfigure}[b]{.45\linewidth}
    \includegraphics[width=\columnwidth]{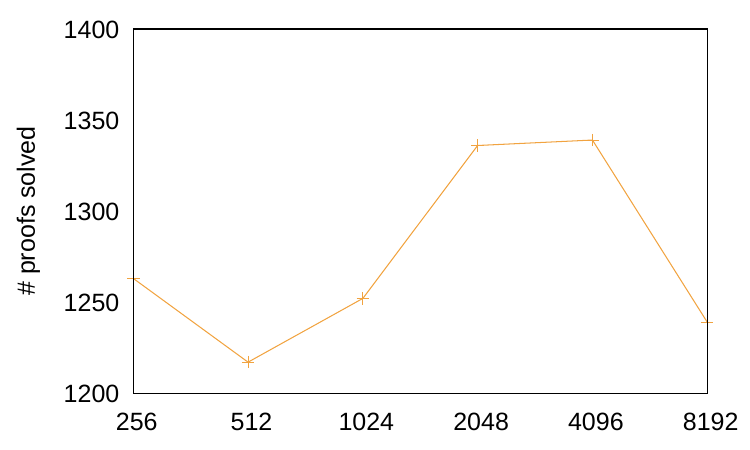}
    \caption{BPE merges}
    \label{fig:bpe-merges-eval}
  \end{subfigure}

  \caption{The impact of different vocabulary thresholds for the various
    categories of identifiers. A smaller threshold means the vocabulary is larger.}
  \label{fig:vocab-size-eval}
\end{figure}

As discussed in Section~\ref{ssec:common-vocab}, each of the
identifier types we add has a vocabulary of the most common
identifiers of that type, giving a fixed encoding of those identifiers
in addition to the subword encoding.
We count the occurrences of the identifiers in the training set to determine
which identifiers occur more than a specified threshold, and then
only include those identifiers in our vocabulary.
For example, if we have a threshold of 100, then all the identifiers
that occur at least 100 times in the training set will be included in
the vocabulary.
That threshold is a hyperparameter that we can vary for each type of
identifier, and it determines the size of the vocabulary.

Figure~\ref{fig:vocab-size-eval} shows the performance impact of
different values of that hyperparameter for different identifiers.
As you can see the performance of various vocabulary sizes for global
definitions, local variables, and type constructors are all fairly
jagged, though they all peak at around 200 occurrences, which we set
as the default in the rest of our experiments.

It is interesting to note that, while the thresholds which produce the
best results are the same for the different identifier categories,
this results in drastically different vocabulary sizes: 427 global
definitions meet the threshold, but only 135 local variables and 26
type constructors do.
This justifies our decision to use a fixed occurrence threshold to pick
vocabulary rather than using the $n$ most common identifiers from each
category.

However, there are signs that our method of picking vocabulary to
index could be improved. Sometimes, adding identifiers with fewer
occurrences, such as the global definitions with between 180 and 200
occurrences, helps; while adding those with more occurrences, such as
the global definitions with between 200 and 220 occurrences, hurts.
This suggests that the number of occurrences does not monotonically
predict the usefulness of indexing a particular identifier, even
though it is the most common approach.
Future systems should investigate new metrics to pick vocabulary for
indexing.
Finally, these experiments indicate that the model is sensitive to
small changes in hyperparameters, similar to how model performance
varied greatly from non-determinism at the hardware level in model
training.

The subword encoding we use also has several hyperparameters which can be
varied; principle among these is the number of byte-pair merges, which
determines the size of the subword vocabulary.
Figure~\ref{fig:bpe-merges-eval} shows the effect of different subword
vocabulary sizes on success rate.
The default byte-pair merge threshold of 4,096 is represented as the
the highest point on the graph.
%% %
%% The number of theorems proven peaks at 4,096 merges and is much less jagged,
%% \todo{I am not sure I agree this graph is less jagged than, say, 11a}
%% indicating that the model is less sensitive to the change in the size of the subword
%% vocabulary compared to a change in the size of an individual identifier vocabulary.

\section{Discussion}
\label{sec:discussion}

We believe that it is prudent to broaden the discourse around machine learning for proofs
to consider not just the tool produced, but also the development processes in building these tools.
It is for this reason that we step back and discuss our experiences,
centering challenges that we encountered in three areas:
the feedback cycle, reproducibility, debugging.

% Check here for useful cites: https://twitter.com/TaliaRinger/status/1509202208084086790
% TODO Ack everyone who responds, also JD, Tom, Louis, folks who responded to later threads, Thomas Latoza, Urs, Stella
% TODO para in intro too

\paragraph{Feedback Cycle}
The feedback cycle for developing \toolname was slow.
Every time we changed an encoding,
we had to retrain the model---a process that took around two days.
Mistakes in the code or in the training parameters
would often not manifest until evaluation,
at which point we would need to retrain once more.
This slow feedback cycle quickly added up,
so that even a small change could take weeks.

In traditional supervised learning training dominates development time, as
evaluating a model just means running it once on the test set.
However, in the context of proof search, evaluation on a large
benchmark set often takes as many or more computational resources as
training, though it is usually more parallelizable across machines.

In the machine-learning literature, techniques have been proposed to
make training
faster~\cite{popel2018training,lepikhin2020gshard,zero,li2022supervision},
which could be directly applied in proof search.
And more tooling like data trackers~\cite{wandb}, data validation, and
static types can help catch bugs sooner, resulting in fewer training
runs needed during development.
Finally, some work in combining multiple models~\cite{First22icse} has
shown an ability to speed up proof search, and other search
optimizations could also shorten that part of the feedback cycle.

%% A recent paper on training transformer models recommends using the smallest
%% model available when evaluating size-independent changes~\cite{popel2018training}.
%% For larger models, recent training optimizations~\cite{lepikhin2020gshard, zero} may help shorten the feedback cycle,
%% as may parallel training across hardware resources.
%% Data-efficient training methods have seen success in other domains like image recognition~\cite{li2022supervision},
%% and may be especially tractable in proof automation,
%% as proof data is so rich that one-shot perfect symbolic generalization is sometimes possible for practical use cases like repair~\cite{ringer2021proof}.
%% Avoiding training queues, when possible, can also help.
%% Experiment trackers like Weights \& Biases~\cite{wandb} may help with preventing mistakes in parameters.
%% Proactive bug prevention measures like testing, data validation,
%% and static types may help cut the feedback cycle short.
%% We hope that more cross-community collaboration
%% will go a long way toward developing new techniques that catch more important bugs before training.

\paragraph{Reproducibility}
As discussed and measured in our evaluation
(Section~\ref{ssec:nondeterminism}), many current learning frameworks
and APIs behave non-deterministically, resulting in non-deterministic
variance in our end-to-end proof results.
Much of the non-determinism we encountered is difficult but possible
to control, when it stems from hardware differences, random seeds, or
OS-level file ordering.
However, even when controlling for those factors and all documented
non-determinism, we found our model training non-deterministically.
During the course of our development, we discovered some PyTorch APIs
which were documented as deterministic behaved non-deterministically;
we reported that bug, and it was marked as
high-priority.~\footnote{https://github.com/pytorch/pytorch/issues/75240}

%% Covert sources of nondeterminism particular to machine learning workflows sometimes
%% made it difficult to reproduce results during development,
%% even after fixing a single random seed and setting PyTorch's determinism flags.
%% For example, %in machine learning, data ordering at training time can affect model performance;
%% we found that our training cluster's operating system returned different default file orders for identical copies of data in different directories.
%% Furthermore, the loss that PyTorch calculated for our model at training time %,
%% %while deterministic on a fixed node of our training cluster,
%% differed across different cluster nodes due to hardware differences.
%% Even fixing a single cluster node, the gradient functions PyTorch computed for us were still nondeterministic.
%% During our investigation, we identified several covert sources
%% of nondeterminism in widely used functions in PyTorch,
%% due to undocumented nondeterminism in the underlying GPU programming interface.
%% We reported this bug to PyTorch, and it was marked as high priority. %\footnote{Link withheld for double-blind review.}

A recent paper found this variance in performance across identical
training runs to be pervasive in an evaluation of six popular neural
networks on three datasets~\cite{variance}.
This paper found that very few of the researchers or practitioners
surveyed in were aware of possible non-determinism in these systems.
We recommend that future researchers using machine-learning for proof
search document the hardware and software used to train, and report
some measure of the variance in their models results.

\paragraph{Debugging}
The debugging of systems that mix machine learning and symbolic
manipulation, such as \toolname, inherits the challenges of both.
Instead of failing to compile or throwing a runtime error, bugs in
\toolname often manifested solely as drops in evaluation numbers.
It was challenging to identify whether these drops were caused by bugs to
begin with, let alone in which part of the system the bug occurred
when there was one. %%  In one case, we lost about three months to
%% tracking down the source of unpredictable regressions in our
%% evaluation numbers.  After months of modifying code, retraining, and
%% reevaluating, we eventually discovered that a script that we had
%% written for data processing was nondeterministically overwriting
%% projects in the training data prior to training.  We were able to fix
%% the script and add data validation prior to training, but this bug was
%% emblematic of the difficulties we faced.

We are unable to find any work on debugging machine learning systems
outside of (potentially very useful) folk knowledge encoded in blog
posts\footnote{\url{http://karpathy.github.io/2019/04/25/recipe/}} and
other informal sources.
Perhaps a more formal exploration of debugging machine learning
systems is warranted.
Both better practices~\cite{popel2018training} and techniques for
improved stability~\cite{liu-etal-2020-understanding} may improve the
debugging experience.
We suspect that improvements to the challenges surrounding the
feedback cycle and reproducibility will be not just helpful for but in
fact \emph{essential to} improving debugging, as many debugging
difficulties are consequences of these challenges.

\paragraph{Other Difficulties}
These were only a few of the difficulties we faced as researchers
applying machine learning to proof search.
These systems are also known to have poor modularity~\cite{techdebt}
(modifying one component can significantly affect the performance of
others); poor explainability~\cite{explain1, explain2, explain3,
  lebese2021proof} (trained models don't lend themselves to high-level
interpretation); and large hardware costs~\cite{hardware-costs}
(expensive hardware is required to train these models, limiting who
can develop them, and often requiring the use of shared clusters which
can slow development).

None of these weaknesses are shared by purely symbolic approaches to
proof tasks such as proof repair~\cite{PumpkinPi}, or first-order
theorem proving~\cite{coqhammer}.
However, current work indicates that tools using these machine
learning models can sometimes overcome limitations that current existing purely
symbolic tools cannot~\cite{First20}, especially when the solution space is large.

\section{Related Work}
\label{sec:related}

% BPE somewhere, and what it has been used for in program and proof synthesis, if anything

We discuss related work in neural proof synthesis,
proof corpora, and neural program synthesis.

\subsection*{Neural Proof Synthesis}

%\begin{figure}[t]
%  \includegraphics[width=\columnwidth]{fig/feature-comparison.jpg}
%  \caption{A comparison of the features of several proof synthesis tools.}
%  \label{fig:feature-comparison}
%\end{figure}

\begin{figure}[t]
  \centering
  \begin{tabular}{lccccc}
 	 &  Proverbot- & ASTactic & TacTok & Passport \\
	 &	 9001	 &		& 		& 		\\
\midrule
Proof search  & \cmark & \cmark & \cmark & \cmark \\
\midrule
Proof state      & \cmark & \cmark & \cmark & \cmark \\
\midrule
Tactic history   & 	---	 & 	---	 & \cmark & \cmark \\
\midrule
Tree-based   & 	---	 & \cmark	 & \cmark & \cmark \\
 term encoder & 	& 		& 		& 		\\
\midrule
Type Constructors   & 	\cmark	 &   ---  &  --- & \cmark \\
\midrule
Global Definitions   & 	\cmark	 &   ---  &  --- & \cmark \\
\midrule
Local Variables   & 	\cmark	 &   ---  &  --- & \cmark \\
\midrule
Paths   & 	---	 &   ---  &  --- & \cmark \\
\midrule
Subwords   & 	---	 &   ---  &  --- & \cmark \\
  \bottomrule
  \end{tabular}
  \vspace{-1ex}
  \caption{A comparison of the features of several proof-synthesis tools.}
  \vspace{-1ex}
  \label{fig:feature-comparison}

\end{figure}

There have been several other neural proof-synthesis tools for the Coq proof assistant.
Figure~\ref{fig:feature-comparison} compares \toolname's features to those of prior work.
Our work directly enriches the TacTok~\cite{First20} proof-synthesis tool for Coq (which is in turn an enrichment of the ASTactic model~\cite{Yang19}),
and evaluates the enriched model on the CoqGym benchmark suite.
TacTok models both proof scripts and proof states to predict tactics.
In doing so, however, it erases all tokens from the AST---effectively erasing all syntactic identifier information,
including path and file names, local variables, theorem names, type names, and type constructor names.
We add these tokens back and explore different design decisions in encoding them,
revealing meaningful information about their contributions,
and improving over TacTok on the CoqGym benchmark suite.
Our insights about syntactic information may provide
ideas for dealing with variables used as arguments to tactics
in future iterations of TacTok.

Other machine learning tools for Coq include Proverbot9001~\cite{proverbot},
Tactician~\cite{tactician}, Gamepad~\cite{gamepad}, and ML4PG~\cite{ml4pg}.
To the best of our knowledge, none of these tools explicitly encode the category a particular identifier belongs to (one of local variable, global definition, or type constructor), none of them encode the path that an identifier comes from, and none of them apply sub-word tokenization.
Our insights may help further improve performance of these tools.

We enrich an existing model to explore the impacts of different design decisions
for including syntactic information.
While the particular architecture of the model we enriched is not the focus of our work,
these design decisions may have different impacts depending on the architecture.
The model we enriched uses a Tree-LSTM architecture;
other models in this space use sequences~\cite{proverbot, holist, tactician},
other tree architectures~\cite{gamepad},
and graph architectures~\cite{christian-gnn},
with the latter showing significant improvement over previous tree architectures.
Models using transformers have also begun to emerge~\cite{gpt-f},
recently showing promising capabilities for benchmarks in Isabelle/HOL~\cite{wu2022memorizing} and Lean~\cite{curriculum}.
Exploring the trade-offs of different encodings of syntactic information
in all of these models may provide interesting insights.

Recent work shows that the decision of whether or not to encode variable names
has a significant impact on the performance of a graph neural network for proof synthesis
in HOL on the HOList benchmark suite~\cite{christian-gnn}.
Our work explores this trade-off at a higher level of granularity,
looking at the impacts of including different kinds of variables and other syntactic information like paths,
and exploring different tokenization decisions and vocabulary sizes.
Running a similar experiment on that tool may also prove enlightening.

\subsection*{Proof Corpora}

A recent study of proof corpora~\cite{Hellendoorn18} applying language models
found high degrees of naturalness in proofs, and discussed implications
for proof engineering tools that could capitalize on that naturalness.
The study also found higher degrees of locality than in other programming languages,
suggesting that cache-based approaches already helpful in neural program synthesis~\cite{localness}
(especially when used in combination with BPE~\cite{big-code})
may prove particularly useful for synthesizing proofs.
Building a cache on top of BPE is a promising path toward further improving our model performance.

The importance of identifiers is also consistent with recent findings from the \textsc{REPLica}
user study of Coq proof engineers~\cite{replica}, which showed a pattern of proof engineers
refactoring the names of definitions in predictable and repetitive ways.
Furthermore, several of the \textsc{REPLica} benchmarks include syntactic changes in proofs
that correspond to semantic changes made alongside them,
which points toward syntactic changes possibly revealing useful semantic information
that a machine learning tool may be able to pick up on.
The \textsc{REPLica} benchmarks may also motivate BPE:
one benchmark, for example, shows a change in a type constructor name,
along with a change of a substring of the name of a broken lemma that referred to that type constructor name
in a way that corresponded to the change.
Exploring the performance of \toolname on those benchmarks may prove interesting.

Nie et al.~\cite{Nie20} developed a model for
auto-formatting Coq code by encoding spacing information in proof
scripts and incorporating techniques from Natural Language Processing.
Their work on Roosterize, a toolchain for generation of lemma
names~\cite{Nie20ijcar, Nie21} leverages both syntactic and semantic
information by combining data from multiple phases of the Coq
compiler---tokens, parse trees, and fully elaborated terms.  Similar
multi-representation approaches may prove an effective means of
encoding syntactic information for proof-synthesis models as well.

Specification-mutation analysis can help demonstrate weak
specifications, when mutating the definitions does not break the
proofs~\cite{Celik19, Jain20}.
iCoq~\cite{Celik17, Celik18}, and its parallelized version
PiCoq~\cite{Palmskog18}, find failing proof scripts in evolving
projects by prioritizing proof scripts affected by a revision. These
tools track fine-grained dependencies between Coq definitions,
propositions, and proof scripts, to narrow down the potentially
affected proof scripts.

\subsection*{Neural Program Synthesis}

Neural proof synthesis is similar to neural program synthesis, but adapted
to the world of proofs.
Neural program synthesis has seen a renaissance of sorts in recent years.
The model beneath Github's Copilot code auto-complete tool---Codex---is
trained on a large corpus of Github projects, and treats all programs
and proofs as text, regardless of the
language~\cite{chen2021evaluating}.
Another work by DeepMind, AlphaCode, solves a similar
task~\cite{alphacode}, as does PaLM-Coder from Google~\cite{palm}.
Work at Google~\cite{google-llm} showed that large language models of
this flavor are promising, but struggle to understand the semantics of
programs.

A recent YouTube video~\cite{joe-talia-hack} explores the applications of Copilot to proofs,
suggesting that even a model trained on raw syntax may suggest helpful hints
for small proofs in repetitive files in the CompCert~\cite{compcert} verified C compiler.
However, it appears to have limited value for larger, more original proofs
with the current data available.

There is a lot we can learn about variable representations and tokenization
decisions in neural program synthesis, some of which may be applicable for proofs.
Recent work~\cite{localness} shows the benefits of a cache-based model
for code completion that exploits locality properties of programs.
More recent work~\cite{big-code} demonstrates the benefits of BPE
tokenization for code completion, especially in combination with cache-based models.
Another recent paper~\cite{fast-code} introduces a framework for evaluating
different design decisions for integrating the structure within identifiers
within a code completion model, and shows similar benefits
for BPE, plus additional benefits from integrating a static analysis to limit the search space.
We find similar benefits to BPE in the context of a neural proof-synthesis model,
and furthermore show the benefits of tagging different kinds of identifiers
and paths differently depending on what kind of information they encode.

Several different models have also been proposed for modeling code,
such as AST-like trees~\cite{tree-models},
long-term language models~\cite{deep-code-model},
and probabilistic grammars~\cite{phog}.
Program synthesis is also widely studied using non-learning based
methods, both from types alone~\cite{complete-completions} and
examples and types~\cite{type-and-example, example-and-type}.

\subsection*{Identifiers in Code Models}

Previous work has been done on providing semantic information for
identifiers in code, outside of the context of proof-synthesis.
The VarCLR paper explored using contrastive learning to learn which
identifiers have similar meanings, in contrast to simply being
related~\cite{varclr}.
It does this by mining variable renamings from GitHub edits, and
enables effective use of general purpose language models.
Another paper~\cite{big-code} explored extensively the
tradeoffs of various techniques for dealing with the large vocabulary
issues that come from modeling identifiers in code.
Several of our design decisions, such as case-sensitivity, and not
attempting to split words based on common conventions, are inspired by
the results of this paper.
This paper also explores the use of subword tokenizing to handle
identifiers in code, and finds it effective.
However, their subword architecture is significantly different than
ours, since it uses a flat sequence model to model unstructured
subword units, while we instead embed a subword model for identifiers
inside of a parsed-tree model of the code structure.

\section{Contributions}
\label{sec:conclusion}

We enriched a model for proof synthesis with three different
identifier encoding mechanisms---category vocabulary indexing, subword
sequence modeling, and path elaboration---to build \toolname.
Each encoding mechanism improved performance of \toolname on the
CoqGym benchmark suite.
Furthermore, we measured the impact of adding information for each
individual category of identifier: global definitions, local
variables, and type constructors, finding that each improved
performance.

These results are consistent with our intuition that identifiers matter for proofs,
that the category of an identifier is useful information, and that drawing connections
between identifiers is useful for proof synthesis.
The final \toolname single-model tool automatically proves 12.7\% of the theorems in CoqGym,
%an improvement of \todo{xx\%} over the previous state of the art, \todo{}.
an improvement of 38\% over the model it enriches---all without
changing the core architecture beyond the encoding of identifiers.
Combining the new models developed in \toolname with the baseline
models, we can automatically prove 17.2\% of the theorems in CoqGym,
an improvement of 45\% over the baseline models combined.
% Talia: we will probably need to say what SOTA is
This intuition and these results will help developers of other tools
for program and proof synthesis in other languages beyond Coq, and
is a fruitful step toward better tools for engineering
robust and reliable formally verified software systems.

%\paragraph{Future Work}
%\todo{Add a future work section, maybe}

\begin{acks}
This work is funded in part by DARPA grant HR0011-22-9-006.
\end{acks}

\bibliography{references}

\appendix

\section{Categories of Identifiers}
\label{sec:appendix}

Before we dove into implementing \toolname, we manually inspected the proof corpora in our training dataset,
walking through proofs and analyzing the kinds of
information needed to make decisions about which tactic to apply
next in a proof.
The choice to include identifiers at all was a product of realizing how much proof engineers rely on naming information to reason about these decisions.
But the choice of \emph{which} identifiers to include was less clear.
Consider, for example, local variables: many common local variable names are used in a variety of contexts which may have little relation with one another.
A variable named \lstinline{x} can carry a totally different meaning than the \lstinline{x} from Figure~\ref{fig:overview-proof-state} in Section~\ref{sec:overview}.
Without empirical evidence, it was unclear whether or not an enriched model could potentially suffer performance degradation from drawing fallacious connections like this.
As a result, experimental data was an important factor in our selection of which identifiers
to include.

Our experiments in Section~\ref{sec:evaluation} show that
all three categories of identifiers help.
In particular, a \toolname model enriched with \emph{any one} of the
three categories of identifiers alone
outperforms a \toolname model with no identifier information.
Furthermore, a \toolname model enriched with \emph{all three} categories of identifiers
at once outperforms a \toolname model enriched with just one category of identifiers,
regardless of the category.

The remainder of this section details each of these three categories---global definitions
(Appendix~\ref{ssec:globals}), local variables (Appendix~\ref{ssec:locals}),
and type constructors (Appendix~\ref{ssec:constructors})---and gives intuition
for why each of them may be useful for a tactic prediction model. Finally, Appendix~\ref{ssec:enrichment} discusses implementation details.

\subsection{Global Definitions}
\label{ssec:globals}
The most straightforward of our categories to include was identifiers
referencing global definitions.
These identifiers refer to objects defined globally directly by the
user, using the keywords \lstinline{Definition}, \lstinline{Theorem},
\lstinline{Inductive}, or one of their variants.
Global definitions are generally either an inductive type name, or a
name given to some Gallina term (function, constant value, etc).
Crucially, since proof objects themselves are terms, theorems are global definitions with their names bound to their proof objects.

In Coq, most code amounts to creating new global definitions, through
a variety of means.
The simplest is by writing the term which corresponds to the name
explicitly, and using a vernacular command to bind it to the name, as
in \lstinline{Definition n := 5.}.
This is commonly how the \lstinline{Definition} keyword is used, both
in defining constant values and in defining functions.
When a definition needs to refer to its own name within its body,
that is done either using a \lstinline{fix} in the term, or using
the special vernacular keyword \lstinline{Fixpoint}, which is
essentially syntactic sugar for the former.

Global definitions can also be defined interactively, using Coq's tactic
system.
For example, the proof script in Figure~\ref{fig:overview-proof-state} specifies a sequence of tactics which produce a Gallina term referred to by its identifier \lstinline{posnatMult_comm}.
In Gallina, this is indistinguishable from a plain definition---in fact, any term in Coq can be defined using tactics, though this is most common for proofs of lemmas and theorems.

Finally, inductive types can be created using Coq's
\lstinline{Inductive} command.
This command creates a new inductive type or type family, given a set
of ``type constructors,'' or ways to build objects of the type.
When complete, this command defines several objects, including the
type itself, its type constructors, and recursion and induction principles
for the type.
Type constructors are explored in more detail in
Appendix~\ref{ssec:constructors}.

Encoding the usage of global definitions in terms is extremely useful
for predicting tactics.
Often, a particular common identifier will signify that certain lemmas
will be useful.
For instance, in the proof context:

\begin{minipage}{\linewidth}
\begin{lstlisting}
  n : nat
  ============================
  le (div2 n) n
\end{lstlisting}
\end{minipage}
the presence of the \lstinline{div2} and \lstinline{le} identifiers
indicates that lemmas involving those operators will be useful; in
fact, the correct next step is to apply a lemma named
\lstinline{div2_decr}, which applies to goals of the form
\lstinline{le (div2 _) _}.
Both \lstinline{div2} and \lstinline{le} identifiers correspond to
global definitions.

\subsection{Local Variables}
\label{ssec:locals}

Besides global definitions, local variables are the most common
type of identifier in Coq terms.
Local variables can be bound to an explicit term, as in a
\lstinline{let} definition, but in many cases (function parameters,
forall bindings, and existential pairs) are given only a type binding.
This is in contrast to global definitions, which are always bound directly to terms.

Encoding local variables is often critical to determining the correct
next step in a proof, or even understanding its basic structure.
Even when the local variable's name isn't particularly informative,
knowing when local variables repeat is often critical.
For example, consider the following proof context (from VST~\cite{vst}):

\begin{minipage}{\linewidth}
\begin{lstlisting}
  n : nat
  ============================
  n >= div2 n + div2 n
\end{lstlisting}
\end{minipage}
If the \lstinline{n} variable weren't the same in all three
occurrences, this goal would be impossible to prove without more
information.
However, because the \lstinline{n} variable is repeated, this goal
holds by the definition of \lstinline{div2}, which is round-down
division by two.

While local variable names often provide useful information, as mentioned above, common names are often overloaded in their usage.
We learned early on that the possibility of performance regression due
to uninformative local variables like \lstinline{x} had concerned the
ASTactic authors, and contributed to their decision not to encode
identifiers.\footnote{\url{https://github.com/princeton-vl/CoqGym/discussions/60}}
However, upon closer inspection of the data we determined that
even single-letter identifier names often carry consistent semantic meaning across proofs.
The identifier names \lstinline{hd} and \lstinline{tl}, for instance, seemed to uniformly refer
to the head and tail of a list; because they carried consistent semantic meaning, these identifiers were treated
similarly within proofs.

Because of these consistencies in naming, we decided to include local
variables.

\subsection{Type Constructors}
\label{ssec:constructors}

Unlike global definitions and local variables, type constructors are
not bound on their own, but are instead defined as part of inductive
type definitions.
As an example of how type constructors are defined, Figure~\ref{fig:option}
shows the definition of the option type.

\begin{figure}[H]
  \begin{lstlisting}
    (* Library Coq, directory Init, file Datatypes.v *)
    Inductive (@\codesimb{option}@) ((@\codesima{A}@) : Type) : Type :=
    | (@\codesimc{Some}@) : A $\rightarrow$ option A
    | (@\codesimc{None}@) : option A
  \end{lstlisting}
  \caption{The polymorphic \lstinline{option} datatype in Coq, found in the
    fully-qualified path \lstinline{Coq.Init.Datatypes}.  Given a type
    parameter \lstinline{A}, an \lstinline{option A} in Coq is one of
    two things: either it is \lstinline{Some a} given an element \lstinline{a} of
    type \lstinline{A}, or it is \lstinline{None}.  For consistency,
    identifiers are highlighted using the same conventions from
    Figure~\ref{fig:overview-definitions}.}
  \label{fig:option}
  \end{figure}

The type definition for \lstinline{option} has two type constructors:
\lstinline{Some}, which creates an \lstinline{option A} for any object
of type \lstinline{A}, and \lstinline{None}, which is a constant value
of type \lstinline{option A} for any \lstinline{A}.
There are many examples of such type constructors in common inductive
types: \lstinline{S} and \lstinline{O} for natural numbers,
\lstinline{cons} and \lstinline{nil} for lists, and others.
Logically, just as type definitions correspond to theorems, type
constructors are analogous to introduction rules for types.
In the option type in Figure~\ref{fig:option}, \lstinline{Some} and
\lstinline{None} encode all possible ways of introducing terms of type
\lstinline{option}.
Because of this, type constructors play a special role in deconstructing
types---in particular, they appear inside match statements, which \emph{act} on the structure of a type by having one branch per type constructor.
Similarly, proofs by induction in Coq \emph{prove} propositions about inductive types by having one case per type constructor.

Knowledge of type constructors can be incredibly useful in determining
the next proof step in a proof.
In the example from Figure \ref{fig:constructors-example}, the goal
states that \lstinline{S (S (n + m))} is even, where \lstinline{m} and
\lstinline{n} are natural numbers.
The context shows \lstinline{(n + m)} is even, but does not include
information about \lstinline{S}.
The knowledge that \lstinline{S} is a successor type constructor of
\lstinline{nat}, and that there exists an \lstinline{ev} type constructor
\lstinline{ev_SS} of type \lstinline{ev n -> ev (S (S n))}, is
necessary to solve the goal.
Here, running the \lstinline{constructor} tactic results in the goal
\lstinline{ev (n + m)}, which matches one of the hypotheses (IH1).

\begin{figure}
  \begin{lstlisting}
  1 subgoal
  m, n : nat
  E1 : ev n
  E2 : ev m
  IH1 : ev (n + m)
  ============================
  ev (S (S (n + m)))
  \end{lstlisting}
  \caption{A mid-proof context from the first volume of the logical
    foundations series~\cite{logfound}}
  \label{fig:constructors-example}
\end{figure}

\subsection{Enrichment Implementation}
\label{ssec:enrichment}

Enriching the data with these three categories of identifiers amounted to
modifying inherited data processing code from TacTok and ASTactic that had
erased all information about those identifiers from the data.
The inherited code had used the SerAPI~\cite{serapi} library to
serialize Coq proof objects (terms) as well as proof states and theorems (types), then processed the serialized ASTs returned by SerAPI to erase all identifier information.
Enriching the data with two of the three categories of identifiers---definition
and local variable names---was a straightforward modification of the
post-processing code.

By contrast, adding type constructor names was a more involved process, as Gallina ASTs do not directly store type constructor names.
Instead, like its parent type theory, the calculus of inductive
constructions~\cite{inductive, coquand}, Coq represents
each type constructor in the AST as a tuple consisting of
the name of its inductive type together with the index of the particular type constructor.

\begin{figure}
\begin{lstlisting}
  (constructor
    (inductive
      (file_path
        (directory_path [Datatypes; Init; Coq])
        (label (@\codesimb{option}@))))
    (int (@\codesimc{1}@)))
\end{lstlisting}
\caption{An unprocessed AST representing a use of the \lstinline{Some}
  type constructor for the \lstinline{option} inductive type from
  Figure~\ref{fig:option}, simplified for the sake of presentation.
  For consistency, identifiers are highlighted using the same
  conventions from Figure~\ref{fig:overview-definitions}, and the index
  \lstinline{1} of the \lstinline{Some} type constructor is highlighted in
  \protect\codesimc{yellow}.  Note that the identifier of the \lstinline{Some}
  type constructor itself is not present.}
\label{fig:constructor}
\end{figure}

Figure~\ref{fig:constructor} shows the AST for \lstinline{Some}, which is the first (type constructors are 1-indexed) type constructor of the \lstinline{option} datatype.
Notably, the AST by default stores the fully-qualified path and name of the
inductive type that the type constructor constructs.
Thus, the only remaining step is to look up the type constructor from the global environment
by passing the fully-qualified name of the inductive type
and the index of the type constructor---here,
\lstinline{Coq.Init.Datatypes.option} and \lstinline{1}---then place
it back into the AST where the index is.

To do this, between parsing and encoding, the \toolname implementation \emph{unparses}
subterms that correspond to type constructor nodes into string representations
of the ASTs of the subterms.
It then feeds those string representations back through SerAPI,
which performs an environment lookup to recover the type constructor name.
As with the other identifiers, \toolname then inserts a child node
containing the identifier into the AST before encoding.

\end{document}